\documentclass[12pt,preprint,natbib]{emulateapj}

\usepackage{txfonts}
\usepackage{gensymb}

\newcommand{\be} {\begin{equation}}

\newcommand{\xmm}{{\em XMM-Newton}}

\newcommand{\bc}{\begin{center}}
\newcommand{\ec}{\end{center}}
\def\ltsima{$\; \buildrel < \over \sim \;$}
\def\lsim{\lower.5ex\hbox{\ltsima}}
\def\loe{\lower.5ex\hbox{\ltsima}}
\def\gtsima{$\; \buildrel > \over \sim \;$}
\def\gsim{\lower.5ex\hbox{\gtsima}}
\def\goe{\lower.5ex\hbox{\gtsima}}

\def\ltsima{$\; \buildrel < \over \sim \;$}
\def\lsim{\lower.5ex\hbox{\ltsima}}
\def\loe{\lower.5ex\hbox{\ltsima}}
\def\gtsima{$\; \buildrel > \over \sim \;$}
\def\gsim{\lower.5ex\hbox{\gtsima}}
\def\goe{\lower.5ex\hbox{\gtsima}}

\def\ergscm2 {erg\,s$^{-1}$cm$^{-2}$}

\def\cm2 {cm$^{-2}$}

\def\mag{Swift~J1834.9--0846}
\def\xmmu{XMMU~J183435.3-084443}

\shortauthors{Torres}
\shorttitle{A rotationally-powered magnetar nebula }

\begin{document}
\title{A rotationally-powered magnetar nebula around Swift J1834.9--0846} 
%further connection between all pulsar classes }

\author{Diego F. Torres\altaffilmark{1,2}}
%\& Yi Wei Bao\altaffilmark{1,3}}
\altaffiltext{1}{Institute of Space Sciences (IEEC-CSIC), Campus UAB, Carrer de Magrans s/n, 08193 Barcelona, Spain}
\altaffiltext{2}{Instituci\'o Catalana de Recerca i Estudis Avan\c{c}ats (ICREA), E-08010 Barcelona, Spain}
%\altaffiltext{3}{Nanjing}

\begin{abstract}

A wind nebula, generating extended X-ray emission, was recently detected surrounding Swift J1834.9-0846. This is  
the first magnetar for which such a wind nebula was found.
Here, we investigate  whether there is a plausible scenario where the pulsar wind nebula (PWN) can be sustained without the need of advocating for 
additional sources of energy other than rotational.
We do this by using a detailed radiative and dynamical code 
that studies the evolution of the nebula and its particle population in time. 
We find that such a scenario indeed exists: Swift J1834.9-0846's nebula can be explained as being rotationally-powered, as all other known PWNe are, 
if it is currently being compressed by the environment. 
The latter introduces several effects, the most important of which is the appearance of adiabatic heating, being increasingly dominant over the escape of particles as reverberation goes by.
The need of reverberation 
naturally explains why this is the only magnetar nebula detected, and provides estimates for Swift 1834.9-0846's age.

\end{abstract}


\section{Introduction }
\label{intro}

J1834.9--0846 was discovered by the Swift X-ray satellite on 2011 August 7, experiencing a short X-ray burst (D'Elia et al. 2011).
% trigger 00458907; 
%
A few hours later, another burst was detected by the {\it Fermi} Gamma-ray Burst Monitor; and a third burst appeared 
days after (on August 29; see Hoversten et al. 2011, Kargaltsev et al. 2012). 
Follow up observations revealed that \mag\ has a spin period $P=2.48$ s and a period derivative (see Table 1) consistent with a dipolar 
magnetic field in the magnetar end of the $P-\dot P $ diagram (Gogus \& Kouvellioutou 2011, Kargaltsev et al. 2012). 
The spin-down power derived from these timing parameters is relatively high  for magnetars, although not unique.
 
 Deep observations in quiescence revealed that \mag\ is surrounded by extended X-ray emission (Kargaltsev et al. 2012, Younes et al. 2012). The inner part of it has a symmetrical shape
 and was  interpreted as a dust-scattering halo. 
 The outer part of the emission has been hypothesized to be a wind nebula. 
 However, the latter was put in question with subsequent observations: Esposito et al. (2013) proposed
 that the spatial, spectral, and time-evolution properties of the X-ray emission surrounding \mag\ were all consistent with a dust-scattering halo due to a single cloud located at a distance of $\sim200$ pc from the pulsar.

Younes et al. (2016) recently reported on new
deep \xmm\ observations of \mag, done 2.5 and 3.1 years after the burst.
They still find an extended emission centered at the magnetar position, asymmetrical, and non-variable (even when comparing with observations dating from 2005).
Younes et al. (2016) provided a discussion of what else, other than a wind nebula, can power this emission. The main contender, i.e.,  scattering of soft X-ray photons by dust, was unfavored due to the constancy of the flux measured between 2011 and 2014 and the hardness of the X-ray spectrum ($\Gamma =1-2)$. 
The latter is at odds with what is expected as a result of a dust scattering of a soft burst emission (when the index was measured to be $\Gamma=3-4$). Thus, this extended emission is different from that found rapidly evolving in time by Esposito et al. (2013).
% see e.g. section 4.1 of younes et al. 2016
% and thus the scattered spectrum would be $\Gamma+2$
In this work, we shall assume these main results of Younes et al. (2016), i.e., that \mag\ is the first nebula to be detected surrounding a magnetar, and try to understand how can it be produced.

Tong (2016) has proposed that the nebula can only be interpreted in the wind braking scenario. Granot et al. (2016) proposed that nebula is powered via a transfer of magnetic power into particle acceleration. 
We come back to these ideas in the discussion. 
However, following the discovery of low-field magnetars (e.g., Rea et al. 2010, 2014), 
and of radio emission from magnetars (e.g., Camilo et al. 2006, 2007; Anderson et al. 2012),
the existence of a rotationally-powered magnetar nebula would only emphasize the connection between all pulsar classes. 
The recent discovery of a magnetar-like burst from the pulsar J1119-6127 (Kennea et al. 2016), which 
is already showing a PWN, further points to the possibility of a PWN origin also for the
radiation from \mag.
Additionally, we have already suggested that magnetar's radio emission can 
be powered by the same physical mechanism responsible for the radio emission in other pulsars (Rea et al. 2012).
%
%We have noted that the magnetars that showed radio pulsed emission were found to have rotational energy loss 
%rates larger than their X-ray luminosity during quiescence.
%
%The low quiescent luminosity of \mag\ in comparison with its spin-down 
%makes it 
%similar to the other radio emitting magnetars. Current upper limits to pulsed radio emission (Esposito et al. 2013)
%are restrictive, but can be compared with hundreds of other pulsars.
%
%In this sense, it would be only reasonable to expect that a nebula can be powered by \mag\ in the same way it is in other pulsars.
%
%In this work, we shall consider this possibility in detail using an advanced PWN radiation code that we have recently developed (Martin et al. 2016).
% 
Because of all of this, we here explore whether the nebula can be rotationally-powered using an advanced radiation/evolution PWN code (Martin et al. 2016). 

The rest of this work is organized as follows. In \S 2 we shall discuss 
the possible relation of \mag\ with the supernova remnant (SNR) W41. 
We shall use this discussion to make initial assumptions on the pulsar's age and distance.
In \S 3 we comment on the spectral energy distribution data that we shall use to constrain the model and in \S4 on the assumptions 
on the environment and on the original explosion that gave origin to \mag. Finally, in \S 5 we present the model in detail, discussing its physical implications in several subsections, and finally comment on our 
conclusions next.

\section{Is \mag\ related with W41?}

The distance to \mag, as with many other pulsars, is unknown. 
Kargaltsev et al. (2012) and Younes et al. (2016), have 
associated to it a plausible distance of 4 kpc.
This estimation is based in part on \mag's location at the geometrical center of the SNR W41, 
for which Tian et al. (2007), and then Leahy \& Tian (2008), estimated a distance of $4 \pm 0.2$ kpc and an age of $\sim 1_{-0.5}^{+1.0} \times 10^5$~yrs.

Tian et al. (2007)  have, however, also considered a possible association of PSR J1833-0827 with W41. 
Hobbs et al. (2005) have also pointed at the 
compatibility of this pulsar's proper motion with a putative W41 association. 
PSR J1833-0827 is located about 10' 
away from the edge of W41, and in addition it
has an estimated distance and a characteristic age that are correspondingly compatible with those of W41. 
Given that the properties of the SNR itself cannot be used to distinguish the magnetar nature of the compact remnant (see the recent work by Martin et al. 2014),
we are uncertain whether \mag\ or PSR J1833-0827  is physically associated with W41.

To complicate things further, there is yet another putative pulsar (no pulsations were measured yet)  that could be associated with W41: XMMU J183435.3-084443 (Mukherjee et al. 2009). 
This object is also close to the center of the remnant 
%(see Younes et al. 2012 for an image showing both this source and \mag\ in relation with W41) 
and it has a faint tail-like emission that could be a PWN
or a dust scattering halo (Misanovic et al. 2011).
With three good candidates for a pulsar-SNR association and no argument of weight other than position (which applies similarly well to \xmmu\ and
even better to PSR J1833-0827)
 the physical connection 
between \mag\ and W41 remains speculative.

In any case, the possibility that \mag\ is located at the vicinity of W41 even if not physically related to it is strengthened by the appearance of 
a dust scattering halo around the magnetar, as observed by Chandra and XMM, following the outburst in September 2011 (Esposito et al. 2013, Younes et al. 2012). 
The dust scattering halo suffered little delay in its flux decay 
compared to \mag, placing a strong constraint on how near the dust should be to the magnetar (about 200 pc, 
Esposito et al. 2013). 
Since the dust estimation is based on CO data, and the Galactic Ring Survey (Jackson et al. 2006) confirms that most
of the molecular material is in the form of a giant molecular cloud nearby W41 rather than a blend of emissions from the
near and far distances (e.g., Albert et al. 2006, Tian et al. 2007), it is reasonable to expect that
the dust, and thus the magnetar, are close to W41. 
%(and all well aligned in projection).
%
Taken this as the main argument (and not just the location of \mag\ at the center of W41), we shall also consider a fiducial 
distance of 4 kpc in what follows. We admit that this assumption should still be taken with care since
up to 
$\pm \sim$1 kpc difference would not introduce much of a change in any of the reasoning that follows.
%
%For instance, Esposito et al. (2013) considered a distance to \mag\ of 5.4 kpc.
%
%We shall come back to this point in the discussion.

Even when we fix the fiducial distance to \mag\ at 4~kpc, there is still the need to consider what is \mag's age.
Of course, this is related to the assumption regarding the physical relationship of 
\mag\ with W41. If they are related, the magnetar age is likely larger than $50$~kyr. 
If they are not, and its localization with respect to W41 happens just by chance, \mag\ can (also) be much younger and compatible with the characteristic age of the pulsar, $
\tau={P}/{2\dot{P}}$ of only 4.9 kyr.
%
%This dichotomy in age was also noted by Younes et al. (2016), although no discussion  was therein provided.
%

If \mag\ would be part of a middle age magnetar/SNR complex (say,  of 60 kyr or older), 
we would need to confront the significant discrepancy with its characteristic age estimation.
In general, for middle-aged or old objects,  $\tau$ overestimates the age. Here, it would be the opposite, albeit 
this situation would not be totally unreasonable a priori.
In the case of PSR B1757-24, for instance, an association with SNR G5.4-1.2 and the consideration of the pulsar's proper motion argue for a real age larger than 39 kyr (and perhaps as large as 170 kyr), with the characteristic age being only 16 kyr ({Gaensler \& Frail 2002}).
For a pulsar of constant moment of inertia, 
the initial spin-down timescale  is (e.g., Gaensler \& Slane 2006)
\begin{equation}
\label{spindownage}
\tau_0=\frac{P_0}{(n-1)\dot{P}_0}=\frac{2\tau_c}{n-1}-t_{age},
\end{equation}
where $n$ is the braking index, and
$P_0$ and $\dot{P}_0$ are the initial period and its first derivative.
Assuming that the angular frequency $\Omega=2\pi/P$ of the pulsar evolves in time as
$\dot{\Omega}=k \Omega^n$
where  $k$ is a constant that depends on the magnetic moment of the pulsar, we find
$n={\Omega \ddot{\Omega}}/{\dot{\Omega}^2} \simeq {P \ddot{P}}/{\dot{P}^2}.$
%
%If the system is a dipole spin-down rotator, the breaking index is exactly 3 and the constant $k$ has the value
%$k={2\mu_\perp^2} / {3Ic^3},$
%where $\mu_\perp$ is the component of the magnetic dipole moment orthogonal to the rotation axis.
%
Only few $n$-values are known, and all but one are less than 3, see Gotthelf (2016). \mag\ would require $n<1.164$ and 
$7.0 \times 10^{42}$ erg s$^{-1}$ of
 initial spin-down power
in order to reconcile $\tau$ with its putative 60 kyr of age.  These parameters would be even more extreme the larger the age.
One can  conclude that the previous formulae are not valid for this (or any) magnetar, e.g, that 
the moment of inertia or the magnetic moment is not constant, and that the current values of $P$ and $\dot P$
are not the result of a pleasant evolution since birth. If so, significant deviations between $\tau$ and the age can occur. Another
possibility  is wind braking (Tong et al. 2013), which can be effective during some epochs of the neutron star life and would also explain
high values of $\dot P$ (and thus reversal of the $\tau > $ age relation). 
Of course, the non-validity of the former formulae can similarly apply to all pulsars.

However, 
since \mag\ is still today close to the center of the remnant, and even within its own nebula, 
is its proper motion directed along the line of sight? In 60 kyr, a velocity of 400 km s$^{-1}$, 
as it has been found typical for pulsars (Hobbs et al. 2005), would have made \mag\ travel 
towards the edge of W41 or even beyond, what is in disagreement with observations. 
%
% 400 km/s * (60000 * 365 * 24 * 60 * 60) * 1E5 (km to cm) / 3.08 E18 (cm to pc) = 24.5 pc
% 1583 arcsec in comparison with <200 arcsec of the extent of the nebula
% at 4900 yrs, the distance travelled would be 30.7/60000*4900=2.0 pc within the nebula (that has 3.8 for 200 arcsec, or 4.8 for 250 arcsec).
%
% the snr has a size of 27 arcmin
% at 4 kpc, this is about 31 pc
%
% the pwn has 2.9 pc at 4 kpc if from figure 1 younes i read its extent as 150 (deviation from 3 sigma from bkg)
%
In fact, being conservative, in order for \mag\ to remain at the central 10\% of the size of W41, it would need to have a traverse velocity smaller than $50$ km s$^{-1}$,  which is a factor of five smaller than the mean of the 2D velocities measured for all pulsars (Hobbs et al. 2005). The larger the age, the larger this discrepancy would be.
Whereas as far as we are aware there is no measurement of the proper motion of \mag, it would be unlikely that its vector direction would allow its current location if physically related to W41.

Thus, on the one hand, because of the scattering halo argument commented above, we shall consider that \mag\ has a  fiducial distance around 4 kpc. On the other hand, and to explore the plausibility of models, we shall consider three cases for the age $\sim60 \% \tau_c$, $\sim\tau_c$, and $\sim1.6\tau_c$ with different braking indices, implicitly unfavoring the association with W41. 
However, we note that whereas there is a constraint on the distance assumption, 
there is no real observational constraint on the age of \mag: It could also be
older if the braking index $n$ is very low, or if the braking is not dominantly magnetic. We just do not know this. 
The assumption of a young age for the magnetar is made with the aim of exploring the plausibility of the usual range of braking indices in a normal evolution of the spin-down, as usually assumed for all other pulsars and their nebulae.

\section{Spectral energy distribution }

%{\it X-rays:}
For the spectral energy distribution (SED) we shall consider the X-ray detection of the magnetar nebula by Younes et al. (2016), which drives this investigation. 
We concurrently consider, also following Younes et al. (2016), that the nebula emission is constant within errors in a period of at least 9 years. 
This is the only part of the SED for which we can be certain it is coming from \mag.

%{\it Radio:}
In the radio band, we take the radio upper limits 
derived from the Very Large Array (VLA) observations at 20 cm (Helfand et al. 2006) and the 1.1 mm Bolocam Galactic Plane Survey (Aguirre et al. 2011)  within a 0.1$^0$ radius around the position of \xmmu\ (see Abramowski et al. 2014). These upper limits are not constraining any of the models we study, and are thus not plotted. 
% and thus valid as well for \mag\ (see Abramowski et al. 2014).
% \mag\ and \xmmu\ are 4.4 arcmin apart.

%{\it GeV:}
At GeV energies, we make use of the upper limits imposed by Li et al. (2016) using 6-years of Fermi-LAT data. 
We have therein provided a dedicated analysis of all magnetars, including \mag. 
For the latter, we removed
the surrounding gamma-ray sources, and modelled out W41, so that our upper limits apply to the 
\mag's nebula directly. We also take into account the detection of W41 itself at higher GeV energies as upper limits for \mag\ emission (e.g, Abramowski et al. 2014).

%{\it TeV:}
The source HESS J1834-087 (Aharonian et al. 2006, Albert et al. 2006) is spatially coincident with \mag. 
%
%This TeV source was also confirmed by MAGIC (Albert et al. 2006). 
%The spectrum and luminosity was found to be compatible with the original measurement. 
%
Dedicated investigations of the TeV source (see Aharonian et al. 2006, Albert et al. 2006, Tian et al. 2007, Li \& Chen 2012, Castro et al. 2013, Abramowski et al. 2014) have all concluded 
that it may be originated in the interaction of particles accelerated in the W41 SNR with molecular clouds in the vicinity. 
Frail et al. (2013) detected OH (1720 MHz) maser line emission, emphasizing the  physical association between the SNR and the molecular material.  
This, together with a hard X-ray spectrum that could originate from synchrotron emission of secondary electrons (Yamazaki et al. 2006),
further sustains a hadronic interpretation. 

PSR J1833-0827 is 20' away from HESS J1834-087 and is thus likely unassociated to it. 
No nebula is known to exist around PSR J1833-0827.
However, if the source around \xmmu\ is a PWN, it could also contribute to the HESS detection;
a magnetar nebula could too. 
Abramowski et al. (2014)  find that the best-fitting model to the HESS excess counts map is that given by a point-like source coincident with the the position of \xmmu\ within errors plus a Gaussian. The latter could represent the hadronically-produced gamma-rays also seen by Fermi; the former point-like TeV source could be produced by the PWN powered by the putative pulsar.
However, the absence of timing parameters for \xmmu\ on which to base the energetics in the model, 
%the disregard of any time dependent evolution of the PWN, and the lack of other critical details 
%e.g., as to what component is generating most of the inverse Compton emission 
would emphasize the care needed to consider these results.
Note too that the SNR-molecular cloud interaction could also generate a rather peaky component at high energies if the cloud complex is formed by compact knots, as seem to be the case (Tian et al. 2007).
This would be similar to the case of IC 443 (Albert et al. 2007, Humensky et al. 2015), also modelled with hadronic interactions (Torres et al. 2010,
Li \& Chen 2012).
Finally, the nearness between \xmmu\ and \mag\ would make them barely distinguishable for an instrument with $\sim 0.1^0$ of angular resolution.

Thus, due to the clear plausibility of other emitters (i.e., a possible \xmmu's nebula, and especially, a SNR-cloud interaction),
we shall consider that the TeV data from HESS J1834-87 represent upper limits 
to the putative TeV emission of the magnetar nebula.
%
%Whatever the latter is, we know for a fact that it should be sub-dominant to the HESS J1834-87 detection. 
%
A less conservative but still sustainable assumption would be to consider that just the TeV point-like source (blue points in figure 7
of Abramowski et al. 2014, about 70\% of the total flux) are themselves the upper limits to the TeV contribution of \mag.

\section{Environment and original explosion}

Tian et al. (2007) concluded that the interstellar medium density (off the clouds) in the vicinity 
of W41 could be as high as 6 cm$^{-3}$, although this would be subject to strong local variations. Given that 
the magnetar seems to be about 200 pc from most of the scattering material (Esposito et al. 2012), we shall consider a range of lower densities (0.1, 0.5, 1, 2, and 3 cm$^{-3}$). 
%This will influence the dynamics and enhance the breemshtralung losses and emission. 

The photon background is unconstrained, but given that the region is dusty and active, it is 
likely above usual 
%(and relatively low in comparison with the location of known TeV PWN) 
Galactic averages. We shall consider, apart the CMB,
two contributions at IR and FIR energies, see Table 1. The TeV emission would function 
as an indirect constraint to the photon energy densities when other parameters of the model (defining the synchrotron part of the spectrum) are defined.

For the explosion of the progenitor star, we shall assume a type II SN. 
Related parameters will be
$E_{sn}$ and $M_{ej}$,   the energy of the supernova (SN) and the total ejected mass during the explosion, respectively. 
We shall consider that the explosion energy is the standard $10^{51}$ erg.
%A larger explosion energy (as considered for newborn millisecond magnetars) could also be accommodated. 

\begin{table}
%\vspace{4cm}
\centering
\scriptsize
\vspace{0.2cm}
\caption{Parameters for the spectral simulations and matching model}
\label{simulation}
\begin{tabular}{lll}
\hline
\multicolumn{3}{l}{ {\bf Measured} } \\
\hline
Period (today) & $P$ & 2.48 s \\
Period derivative (today) & $\dot P$  &  $7.96 \times 10^{-12}$ s s$^{-1}$ \\
Nebula radius (today, at $d$) & $R$ & 2--4 pc  \\
\hline
\multicolumn{3}{l}{  {\bf Computed from $P$ and $\dot P$, assuming  $n,t_{age}$}  } \\
\hline
Characteristic age (today) & $\tau$ $= P / [2 \dot P$] &  $4.9 \times 10^3$ yr \\
% tau used = 4950 yrs
Spin-down luminosity (today) & $L_{sd}= 4 \pi I \dot P / P^3$ & $2.1 \times 10^{34}$ erg s$^{-1}$ \\ 
Dip. magnetic field (equator, today) & $B_{dip} = 3.2 \times 10^{19} (P \dot P)^{1/2}$  & $ 1.4 \times 10^{14}$ G\\
Initial spin down age & $\tau_0 $  & [depends on $n,t_{age}$] \\
Initial spin down luminosity & $L_0$  & [depends on $n,t_{age}$]  \\
%Spin_down_luminosity_now*(1+Age/Initial_spin_down_age)**
%#                                  ((Braking_index+1)/(Braking_index-1))
\hline
\multicolumn{3}{l}{  {\bf Magnetar assumptions} }\\
\hline
Distance & $d$ & 4 kpc\\
Real age & $t_{age}$ & $\ll$50 kyr \\
Braking index & $n$ & [2--3] \\
\hline
\multicolumn{3}{l}{ {\bf Supernova Explosion and environment} } \\
\hline
%PWN adiabatic coefficient & $\gamma_{pwn}$ & 4/3\\
%SNR adiabatic coefficient &$\gamma_{snr}$ & 5/3\\
%Index of the SNR density power law & $\omega$ & 9\\
Energy of the Supernova & $E_{sn}$ & $ 10^{51}$ erg\\
Ejected mass & $M_{ej}$  & [7--13] M$_{\odot}$\\
ISM density & $\rho$ & [0.1--3] cm$^{-3}$\\
%CMB temperature & $T_{cmb}$  & 2.7 K \\
%CMB energy density & $w_{cmb}$ (eV cm$^{-3}$) & 0.26 eV cm$^{-3}$\\
%FIR temperature & ($T_{fir} = 25 K)\\
%NIR temperature & ($T_{nir}$  = 3000 K) \\
\hline
\hline
\multicolumn{3}{l}{  {\bf Specific spectral model (see text)} }\\
\hline
Real age & $t_{age}$ & 7.97 kyr \\
Braking index & $n$ & 2.2 \\
Initial spin down age & $\tau_0 $&  280 yrs\\
Initial spin down luminosity & $L_0$ & $1.74 \times 10^{38}$ erg s$^{-1}$  \\
%Minimum energy at injection & $\gamma_{min}$ & 1\\
Ejected mass & $M_{ej}$  & 11.3 M$_{\odot}$\\
ISM density & $\rho$ & 0.5 cm$^{-3}$\\
Energy break at injection & $\gamma_b$ & $10^7$\\
Low energy index at injection & $\alpha_l$ & 1.0\\
High energy index at injection & $\alpha_h$ & 2.1\\
Containment factor ($<1$) & $\epsilon$ & 0.6\\
Magnetic fraction ($<1$) & $\eta$ & 0.045\\
Nebular magnetic field & $B$ & 4.8 $\mu$ G\\
FIR energy density  ($T_{fir}$ = 25 K) & $w_{fir}$  & 0.5 eV cm$^{-3}$\\
NIR energy density ($T_{nir}$  = 3000 K) & $w_{nir}$  & 1 eV cm$^{-3}$\\
\hline
\hline
\end{tabular}
\end{table}

%note that using the evolution of P(t) in time, see e.g, eq. 4 of granot et al 2016
%the P_0 is ~170 ms.        

\section{Model and results}

%%%%%%%%%%%%%%%%%%%%%%%%%%%%%%%%%%%
%\begin{center}
%\begin{figure*}
%\centering
%\includegraphics[scale=0.45]{colourmap_radii_3000.eps} 
%\includegraphics[scale=0.45]{colourmap_radii_4935.eps}
%\includegraphics[scale=0.45]{colourmap_radii_7900.eps}
%\caption{
%Left: 
%Radius of the PWN for different ejecta mass and densities for an estimated age of 3000 years. The braking index is assumed as 2.5. 
%A circle implies that the PWN is in free expansion. Middle: The same but for an age of 4935 years. A square (triange) implies that the PWN is in compression (Sedov) phase.
%Right: The same for an age of 7900 years and braking index 2.2.
%The instantaneous sharing of spin-down power into magnetic energy is 0.03 in all cases.}
%\label{1}
%\end{figure*}
%\end{center}
%%%%%%%%%%%%%%%%%%%%%%%%%%%%%%%%%%

%
We consider that the magnetar is constantly injecting high energy particles similarly to what happens in any other rotationally-powered PWN.
We use the model described in detail in Martin, Torres \& Pedaletti (2016) to study the radiation of the magnetar nebula.
This model computes the evolution in time of the pair distribution within the PWN, subject to 
synchrotron, inverse Compton, and Bremsstrahlung interactions, adiabatic losses, and accounting for escaping particles. 
Critical formulae are shown in the Appendix.
Full expressions for the radiative losses can be found in Martin, Torres, \& Rea (2012).
All time dependences
are included both at the level of spin-down, injection, and losses. 
The magnetic field varies in time as a result of the balance between the instantaneous injection (the fraction of spin-down that goes to power the magnetic field, $\eta$, which is assumed constant in time) and the adiabatic losses of the field due to the expansion of the PWN. The radiative losses of the particle population are taken into account also when determining the inner pressure of the nebula. The injection function for pairs is assumed as a broken power law. The radius of the PWN is computed taken into account reverberation processes when they happen, according to age, progenitor explosion, medium density, and pressure. After reverberation, a Sedov expansion is activated when 
the PWN pressure reaches that of the SNR's Sedov solution. 
No additional source of energy, other than rotational, is assumed in the model.

\subsection{Size constraint}

Assuming 
that a  $\sim $3$\sigma$ deviation from the background level represents 
the radius of the nebula, it is $\sim$2.9 pc (equivalent of 150 arcsec at a distance of 4 kpc, following Fig. 1 of Younes et al. 2016).
Given that anisotropies are not considered,
a radius between 2 and 4 pc would then be an acceptable outcome of models: The aim of this first exploration is not to match the spectra,
but rather to find out in which regions of the phase space, an acceptable radius could arise. 

At different putative ages, the PWN evolution should be such that the $\sim$2.9 pc-radius is observed today. 
This necessarily implies
differences in models, e.g., regarding evolutionary stage, medium density, ejected mass, and magnetic field; in case the size constraint
can be accommodated at all.
The larger the density, for instance, the faster the nebula will be in reverberation, or even pass this stage to slowly expand in the Sedov phase.
Let us now briefly consider the three putative ages noted in \S 2 with regards of the nebula radius.

 {\it Age $\sim60 \% \tau_c$:} Here we shall assume that the really young, 3000 years-old magnetar has a braking index $n=2.5$, in which case the spin-down age is larger than the pulsar's, 
%
%> b := 2.5; age := 3000; spindownage := (2*4940)/(b-1)-age;
%                              2.5
%                              3000
%                          3586.6
%
or $n=3$, in which case, the pulsar's spin-down age is about 2/3 of its real age. 
In neither of these cases, we are able to obtain a satisfying solution: the PWN is not energetic enough to inflate a bubble of ~3 pc.
The nebula would be in free expansion (see the range of other explored parameters in the first panels of Table 1), but given the short age, the size of the nebula would be smaller than observed.
%
%Examples of these results for ($n=2.5$) are shown in Fig. 1, for a range of ejecta masses and densities.
%

{\it Age $\sim\tau_c$:}
We have also explored the same two values of braking indices (2.5 and 3) for an age similar to $\tau_c$. 
In this case, for medium densities  larger than 1 cm$^{-3}$, the PWN would be at or have past reverberation
for essentially any value of the ejected mass. This implies a significant reduction in size.
The compression is not fully compensated when the Sedov expansion continues, since 
the nebula grows at a slower pace than in the initial free expansion phase.
%
%Note that when the PWN is in free expansion, the density of the ISM plays no role
%in defining the PWN radius.
For smaller medium densities, we find nebulae in free expansion, but they do not reach a size of about 2.9 pc either, similarly to the case 
above.

 {\it Age $\sim1.6 \tau_c$:}
For an age of about 7900 years, we have explored braking indices equal to 1.7 and 2.2, { so that we continue to use Eq. 1}.
%and show the latter in Fig. 1. 
%
Here again, the three evolutionary stages appear, although for most of the values of the medium density but the smallest ones explored (e.g., 0.1, 0.5 cm$^{-3}$) the nebula would already be past compression or at this stage, and a significant reduction in size is found as a result thereof.
However, the radius of the nebula can be matched at this age
for a range of models having
ejecta mass and medium density within reasonable values.

One can conclude that a young magnetar could perhaps  accommodate 
the X-ray results in a limited range of ages only: For the smaller values of ages considered, the pulsar would be too young to be free-expanding a rotationally-powered nebula up to the size detected;
for larger values of age, 
the PWN expansion would have been already stopped by the medium and even when re-expanding, its size would be smaller than detected. 
Solutions preferred from the perspective of the nebula radius are then those having an age around $1.6 \tau_c$,
 at the end 
of the free expansion phase or the beginning of the compression phase, where the nebula has not yet time to be compressed too much
by the reverberation process. Now we must also see whether the predicted 
spectrum and the X-ray luminosity can match the observations.

\begin{center}
\begin{figure}
\centering\includegraphics[scale=0.35, angle=-90]{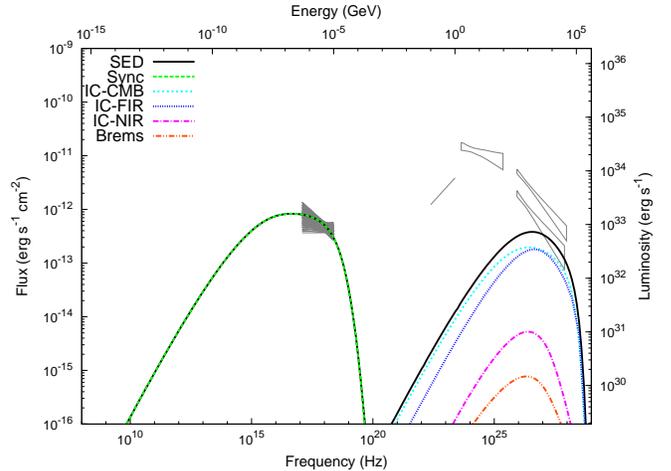}
\caption{Matching solution with 7970 yrs of age for a rotationally-powered magnetar nebula at the start of the reverberation phase. See text for the SED-data references, and for discussion.}
\label{7900-fit-SED}
\end{figure}
\end{center}

\subsection{Example of a spectrally-matching model}

We have fixed the ejecta mass and medium density to 11.3~M$_\odot$ and 0.5 cm$^{-3}$, respectively, for further exploration; and considered the standard explosion energy.
Within this scenario, we find that the spectrum can be matched with reasonable choices of other parameters related to injection and the magnetic field, which result to be rather similar to other rotationally-powered PWNe.
An example is given in Fig. \ref{7900-fit-SED},
which shows the predicted spectrum at the pulsar's age, resulting in 7970 yrs,
in comparison with X-ray data and multi-frequency constraints. 
The parameters used are given in the last panel of Table 1. 
Based on this result, the magnetar nebula could be rotationally-powered. 

\begin{center}
\begin{figure}
\centering
\includegraphics[scale=0.65]{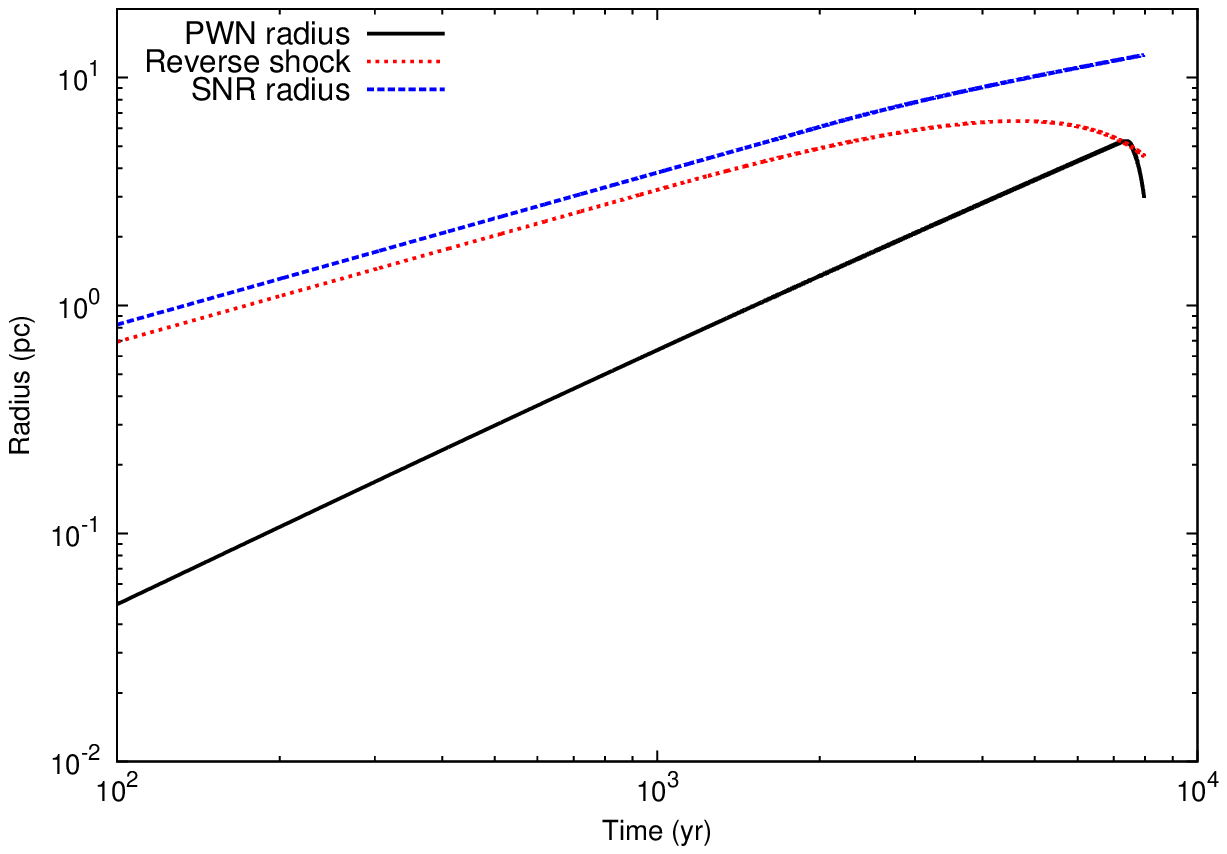} 
\includegraphics[scale=0.65]{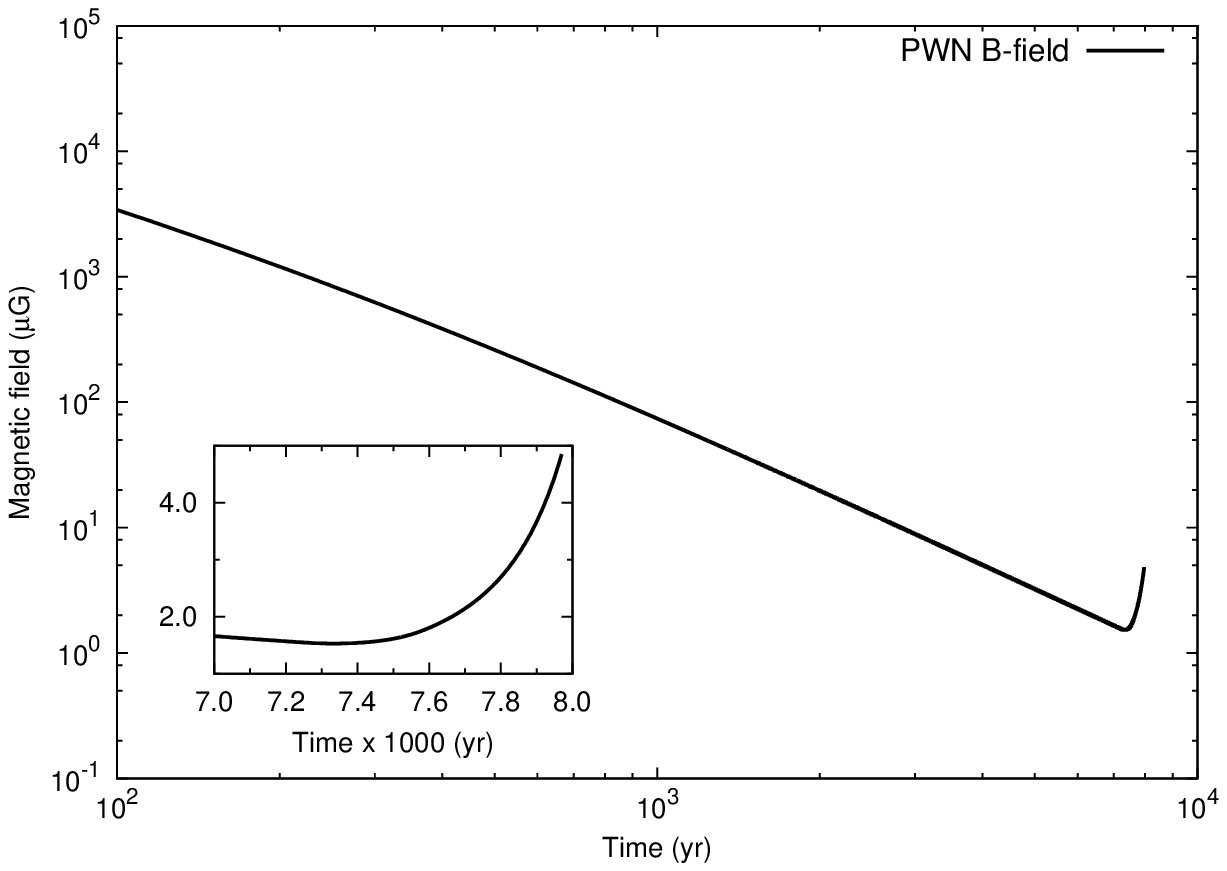}
\includegraphics[scale=0.65]{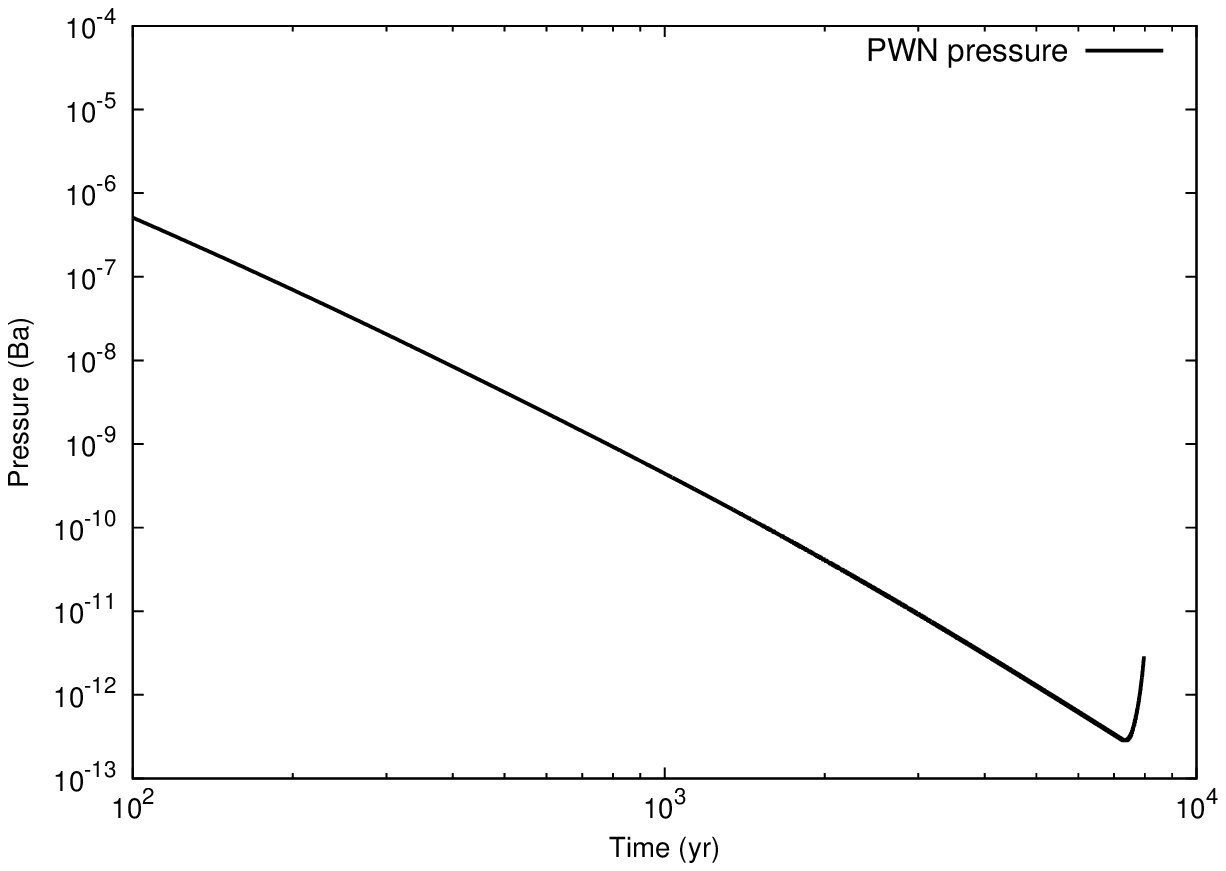}
\caption{Details of the spectral-matching model for the \mag's nebula presented in Fig.  \ref{7900-fit-SED}. The panels show the time evolution 
of the nebula relevant radii, magnetic field, and pressure. }
\vspace{0.2cm}
\label{7900-fit}
\end{figure}
\end{center}

We shall now analyze the solution in detail, and for this we shall also make use of the characterization
of the model given in Fig. \ref{7900-fit}. 
We shall also compare this solution with those obtained for other nebulae.  
In particular, we shall use 
the results obtained for CTA 1, where the theoretical model used was the same (see Martin et al. 2016), and 
the studies 
of nebulae presented earlier by Zhang, Cheng \& Fang (2010), Tanaka \& Takahara (2010), Bucciantini et al. (2011), Vorster et al. (2013), 
and Torres et al. (2013a, 2014).
Apart from significant differences among themselves, which sometimes caveat a direct comparison,
the models used in most of these latter cases do not account for reverberation. In fact, models including reverberation are lacking in the literature for { most} nebulae. Still, this comparison is 
useful for context, in the understanding that the low-age and powerful nebulae 
studied earlier are most of them free-expanding.

%\subsection{Characterization at the current age}

%%%%% RADII

The first panel of Fig. \ref{7900-fit} shows the evolution in time of the relevant radii in the model. 
The nebula has started to compress about 400 yrs ago, when the reverse shock got to the PWN shell. 
Today the modelled nebula has a radius of 2.9 pc, compatible with observations.
The SNR radius 
is at about 10 pc, what would place it near to, and for the most part superposed with W41. This 
would make its detection, as well as its differentiation with W41, extremely difficult.
%

%%%%% MAGNETIC FIELD, magnetization, pressure

The matching spectrum shown in Fig. \ref{7900-fit-SED} is obtained with an instantaneous sharing of spin-down power into magnetic energy $\eta=0.045$. 
This low percentage for the magnetic field energization is rather similar to all other nebulae modelled, 
making this one too a strongly particle-dominated system.
Today's magnetic field in the nebula (see second panel of Fig. \ref{7900-fit}) is 4.8~$\mu$G,  and make this case similar to G292.2-0.5, HESS J1336-645, or CTA 1. The value of $\eta$ and the field puts the nebula far from equipartition, what is also the case for all other systems studied (see 
references quoted in \S 5.2).
As shown in the inset of the second panel, 
the magnetic field is currently being enhanced by the compression process. 
The same happens with the nebula pressure (see third panel of Fig. \ref{7900-fit}).

%%%%% ELECTRONS

The low spin-down power reflects on the dim spectral energy distribution shown in Fig. \ref{7900-fit-SED}
but also on the electron spectrum (discussed in the next sections), which maximum is about two orders of magnitude smaller than that obtained for more luminous nebulae (e.g., G54.1+0.3, G0.9+0.1, or G21.5-0.9). This is again comparable to the cases of G292.2-0.5, or CTA 1 and also of  Kes~75.

%%%%% FLATNESS of X-ray SPECTRA // MAXIMUM ENERGY at injection // 

%The X-ray spectrum is rather flat.
%
%This implies accommodating a correspondingly flat energy index at injection
%with a high energy break (i.e., where the injected broken power-law changes slope), or a high containment factor, or both. 
%
%The containment factor is the ratio between the Larmor radius of the pairs and the termination shock radius (see e.g., de Jager and Djannati-Atai 2009). 
%The containment factor influence specially the high end of the X-ray spectrum.
%
In the matching model presented in Fig. \ref{7900-fit-SED}, 
%we solve the constraint introduced by the X-ray spectrum flatness by admitting 
we have a Lorentz factor for the electron break at injection of $10^{7}$ and a containment factor of 0.6.\footnote{The containment factor is the ratio between the Larmor radius of the pairs and the termination shock radius (see e.g., de Jager and Djannati-Atai 2009). }
The break appears to be at the high end of the previously-studied nebulae. 
It is comparable (a factor of 2 higher) to the break fitted in the case of G292.2-0.5,
and about a factor of 10 higher than that found for most other nebulae. However, the interplay with dependences on other spectral parameters 
can reduce the break energy, and we should consider  that reverberation may also impact onto the break value. We discuss both of these effects  below.
In any case, 
the high Lorentz factor of the break implies that particles are accelerated up to TeV energies. 
This is needed since the low value of magnetic field would require such high-energy particles to emit X-ray photons:
The electron energy $E$ needed to produce a synchrotron photon with keV energy is 
(e.g., De Jager and Djannati-Atai 2009)
\begin{equation}
E \sim 220 E_{{\rm keV} }^{1/2} B_{\mu {\rm G}}^{-1/2} {\rm TeV}.
\end{equation}
Still,  the magnetar is not a pevatron at any moment of its history.
% see fifth panel of Fig. \ref{7900-fit}.
% see figure a-plot_gmax.eps
% 
The maximum energy at injection  increases up to a Lorentz factor of less than $\sim6 \times 10^9$ in the first few hundred years of the magnetar's life, and then decreases to $\sim2 \times 10^{8}$ today. This change of behavior  appears due to the influence of two limiting constraints, the synchrotron limit at early times, and 
the confinement of the
particles inside the termination shock of the PWN, thereafter (see Appendix).

The values of the spectral indices at injection are comparable to those obtained for other nebulae, including those associated with more powerful 
pulsars. The lack of radio data gives an extra degree of freedom to these values.

The different components contributing to the spectrum at high energies are noted in Fig. \ref{7900-fit-SED}, and the energy densities of the NIR and FIR photon background 
are chosen such that their contribution makes the spectrum still compatible with the most restrictive of the TeV constraints. Inverse Compton against the CMB photons dominates the spectrum at high energies, but it is not far from the FIR contribution for the chosen energy density. The self-synchrotron contribution is out of the plot in this scale.  If this model for \mag\ 
realizes in nature, the nebula could be particularly contributing to the high energy end of the TeV spectrum. 

%The sixth panel of Fig. \ref{7900-fit} gives account of the timescales of the different losses today (including the effective timescale). 
Considering particle losses (or gains), the values for the timescales   are 
\begin{equation}
t_i = \frac{ \gamma }{ \dot\gamma_i (\gamma,t) }, 
\end{equation}
with $\gamma$ being the particle's Lorentz factor, the sub-index $i$ representing each of the processes, and
$ \dot\gamma_i (\gamma,t) $ standing for the particle's energy change.
At low energies, the adiabatic timescale, 
\begin{equation}
t_{ad} \sim \frac{ R(t) }{ v(t) },
\end{equation}
is the smallest (which is the usual case). Since the nebula is in reveberation, the adiabatic timescale along compression actually represents a characteristic time for gains (and not losses) in the particle's energy. 
We discuss this situation in detail in \S 5.4.
At higher energies, and due to the small magnetic field, 
the escape of particles, which we described via Bohm diffusion,  
\begin{equation}
t_{\rm Bohm} \sim \frac{q B(t) R(t)^2 }{ 2 m_e c^3 \gamma },
%       escape=E*b*r*r/(2.d0*ME*C*C*C*gamma)
\end{equation}
dominates over synchrotron losses; { ($q,m_e$) are the electron charge and mass, respectively.} Synchrotron losses are defined as 
\begin{equation}
t_{\rm Sync} \sim \frac{ \gamma }{ \frac{4}{3}\frac{\sigma_T}{m_e c}U_B(t)\gamma^2 },
\end{equation}
%with $\dot{\gamma}_{sync}(\gamma,t)=\frac{4}{3}\frac{\sigma_T}{m_e c}U_B(t)\gamma^2 ,$
where $\sigma_T=(8\pi/3)r^2_0$ is the Thomson cross section, $r_0$ is the electron classical radius, and $U_B(t)=B^2(t)/8\pi$ is the energy density of the magnetic field. 

\subsection{Time evolution}

\begin{center}
\begin{figure}
\centering
\includegraphics[scale=0.70]{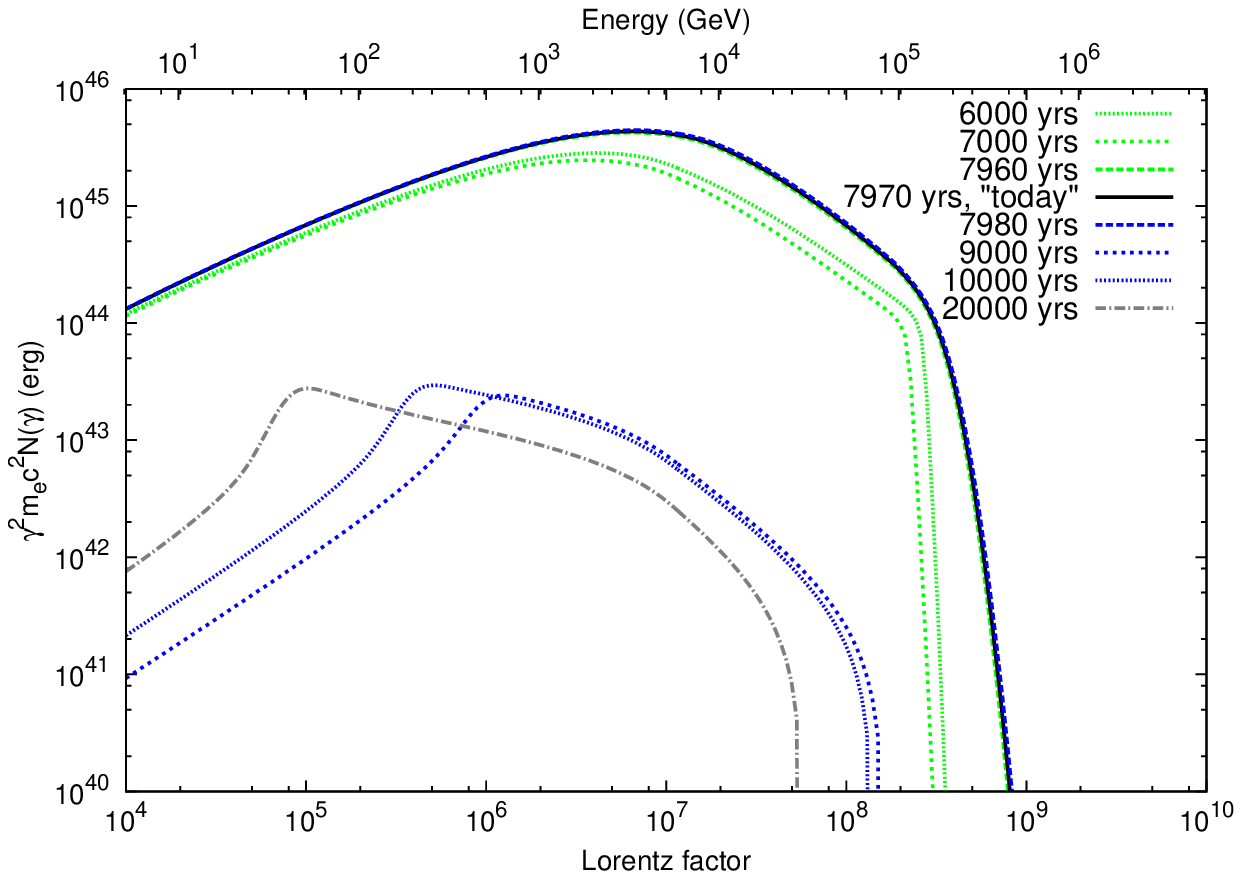} 
\includegraphics[scale=0.70]{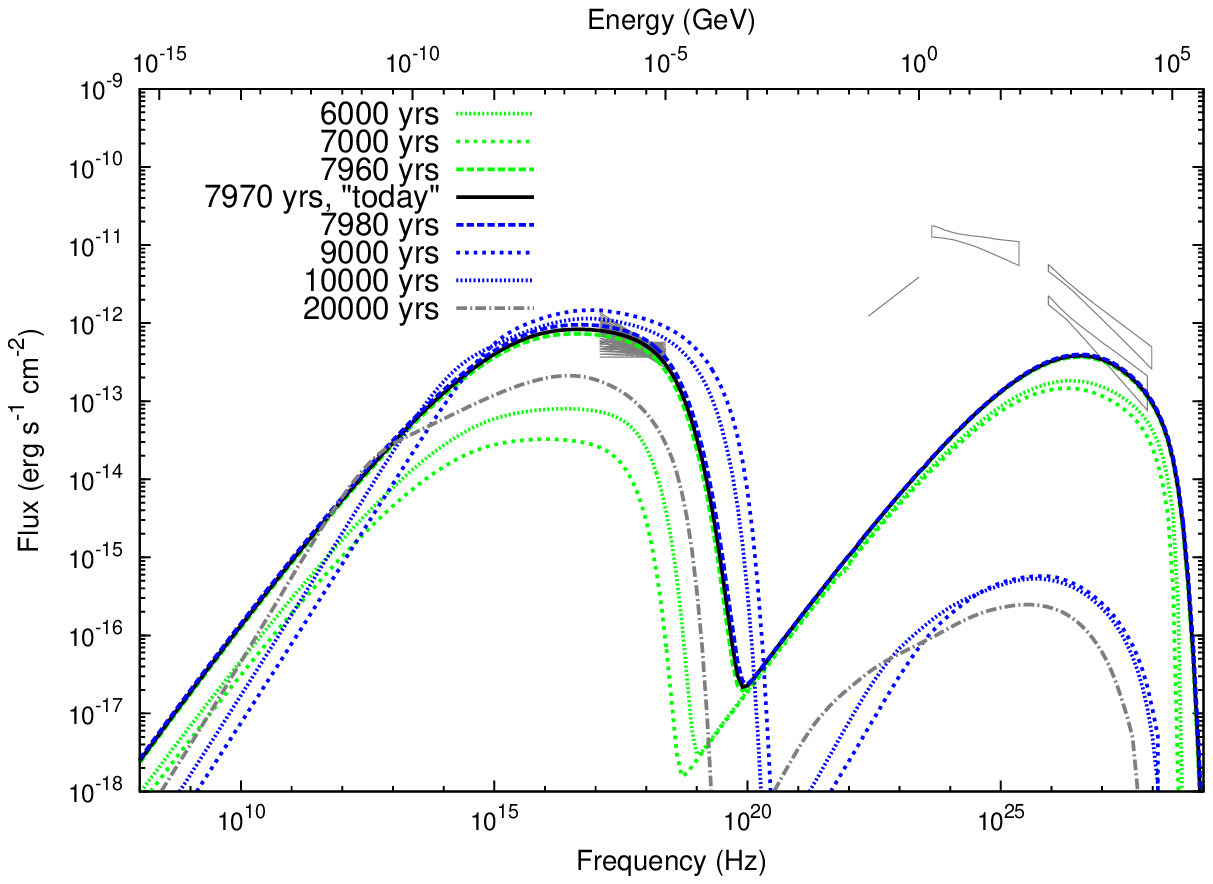} 
\caption{Time evolution of the matching solution discussed in Fig. \ref{7900-fit}: electrons and SED. }
\vspace{0.2cm}
\label{7900-TE}
\end{figure}
\end{center}

Fig. \ref{7900-TE} shows the time evolution of the electron spectrum and the 
SED for the matching solution presented in Fig. \ref{7900-fit-SED}. 
The model predicts no significant  change in the X-ray output of the nebula in a timescale of a decade around the age today.
This is in agreement with observations too, and predicts that the nebula luminosity in X-rays will also remain rather stable  in the following years. 
However, this is not the case forever, and  our descendants  should see a significant time evolution of the SED.
This evolution is not linear, since the reverberation process is rather steep. 
The system will be more luminous first, until most of the high-energy electron population are burned off and re-expansion starts. 
In this solution, the bounce will happen at about 8370 yrs. 
The electron distribution will significantly decrease in the re-expansion of the nebula (see the population at 9000 yrs and beyond). 
However,  at the beginning of the re-expansion, the concurrent high-value of
the magnetic field due to the earlier compression (it is still at about 200 $\mu$G at 9000 yrs, for instance) will make for a sustainable synchrotron power. 
This explains why the level of the X-ray part of the SED is maintained or even increases up to this time, despite the significant reduction in the number of pairs. 
Such magnetic effect does not apply
to the inverse Compton yield, which will then decrease all along.
When pairs are burned off at the start of the Sedov phase, while the high-energy particles are slowly building up only by injection, the synchrotron luminosity will finally decrease (see the SED at 20000 yrs). 
The radius that this future nebula will have is smaller than 1 pc, having been affected by the strong compression process, which goes almost unimpeded due to the relatively low pulsar power.

It is interesting to note too what has happened before reaching the current age. 
Fig. \ref{7900-TE} compares the current SED and electron spectrum with their values 
at 6000 and 7000 yrs. 
At both of these ages along the nebula evolution, the pulsar was still free expanding. 
The larger the age (at 7000 vs. 6000 yrs), the PWN was larger (5.0 vs 4.3 pc), the magnetic field was smaller (1.65 vs 2.25 $\mu$G), and 
more particles were affected by losses, for which the timescales started to be of comparable magnitude to the age (see next). 
Thus, the electron population is larger at 6000 yrs than at 7000 yrs, and the SED is correspondingly more luminous.
What happens next, between 7000 yrs and today, depends strongly on the reverberation process, and constitutes
the reason why the nebula can actually be observed.

\subsection{The impact of the dynamical evolution of the nebula}

To understand why the impact of reverberation on the ability of the model to match the data is 
critical, we now consider models without 
reverberation. In these models, the PWN is considered to be free-expanding all the way up to the same assumed, current age.
All parameters are as those given in Table 1.

\begin{center}
\begin{figure}
\centering
\includegraphics[scale=0.62]{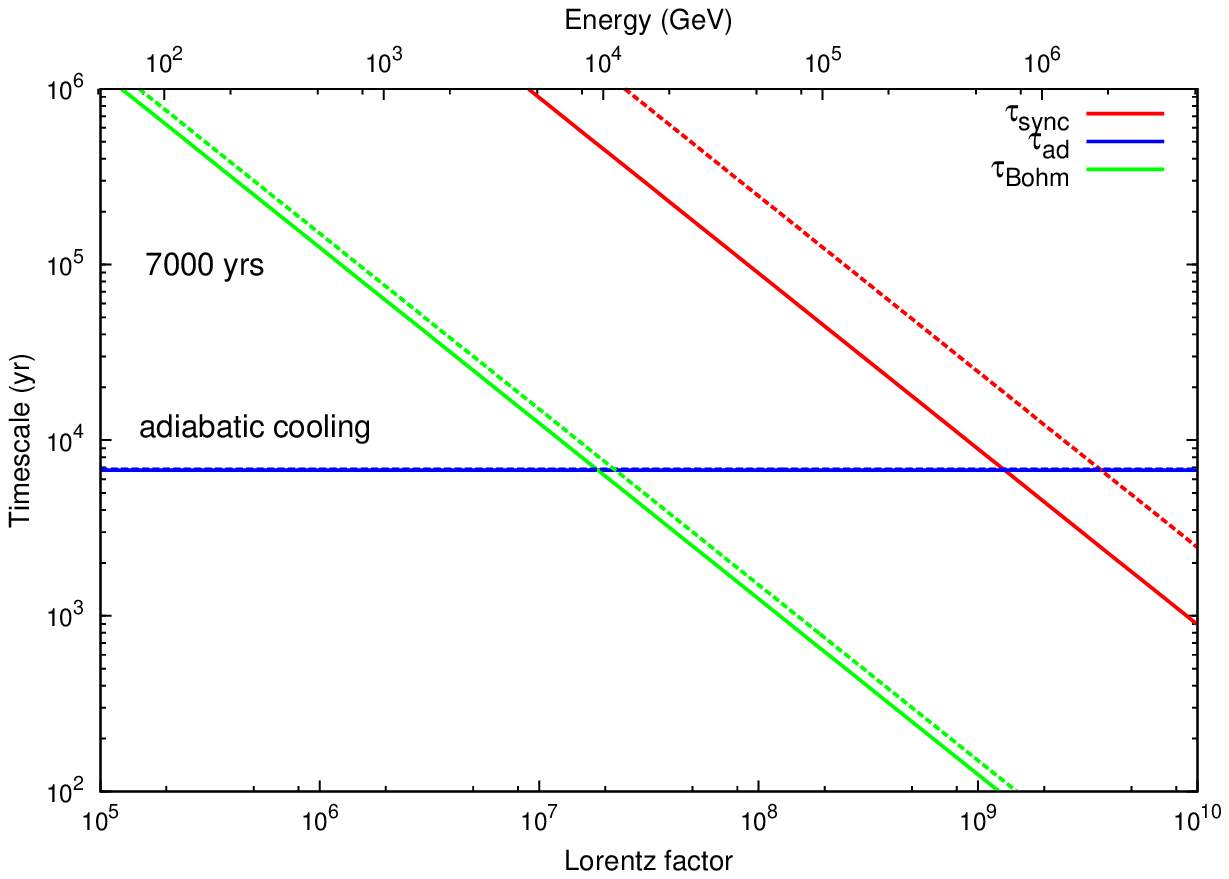}
\includegraphics[scale=0.62]{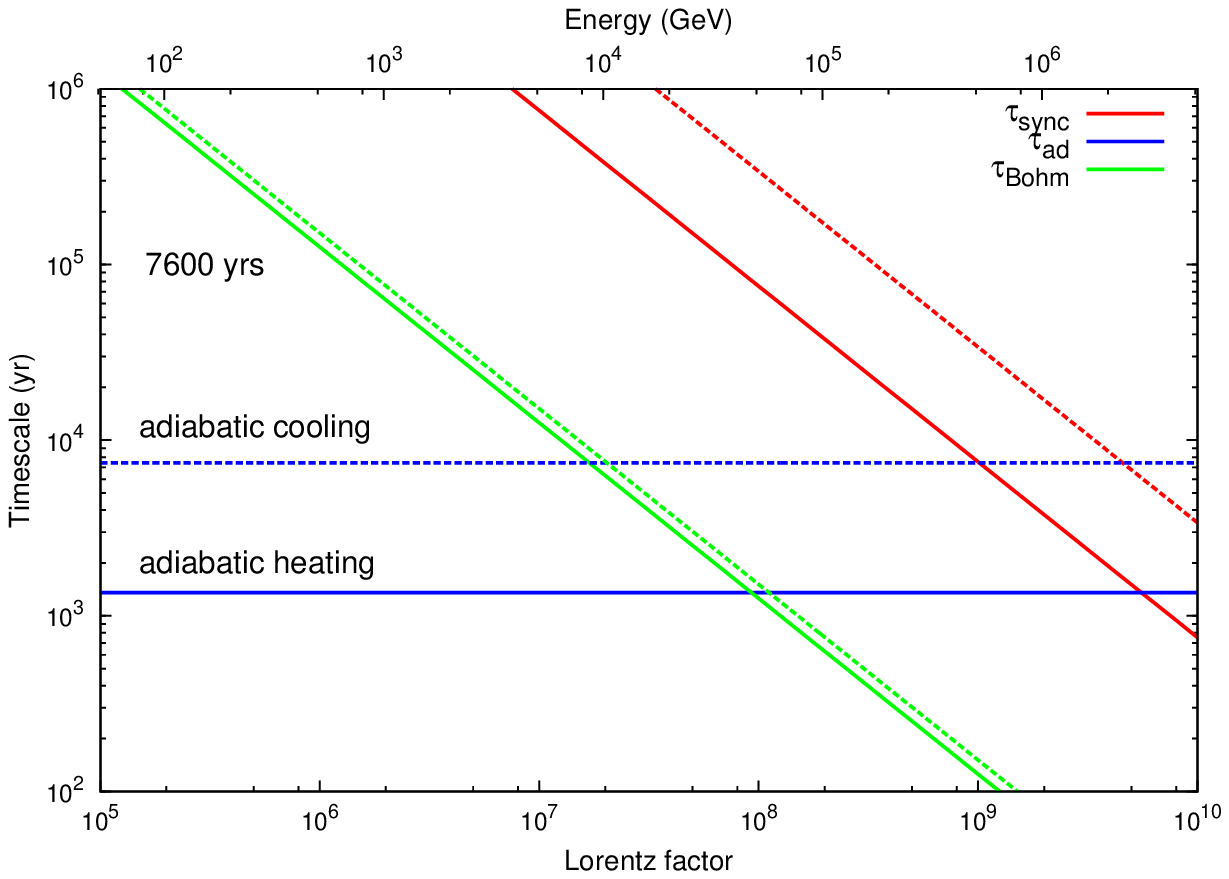}
\includegraphics[scale=0.62]{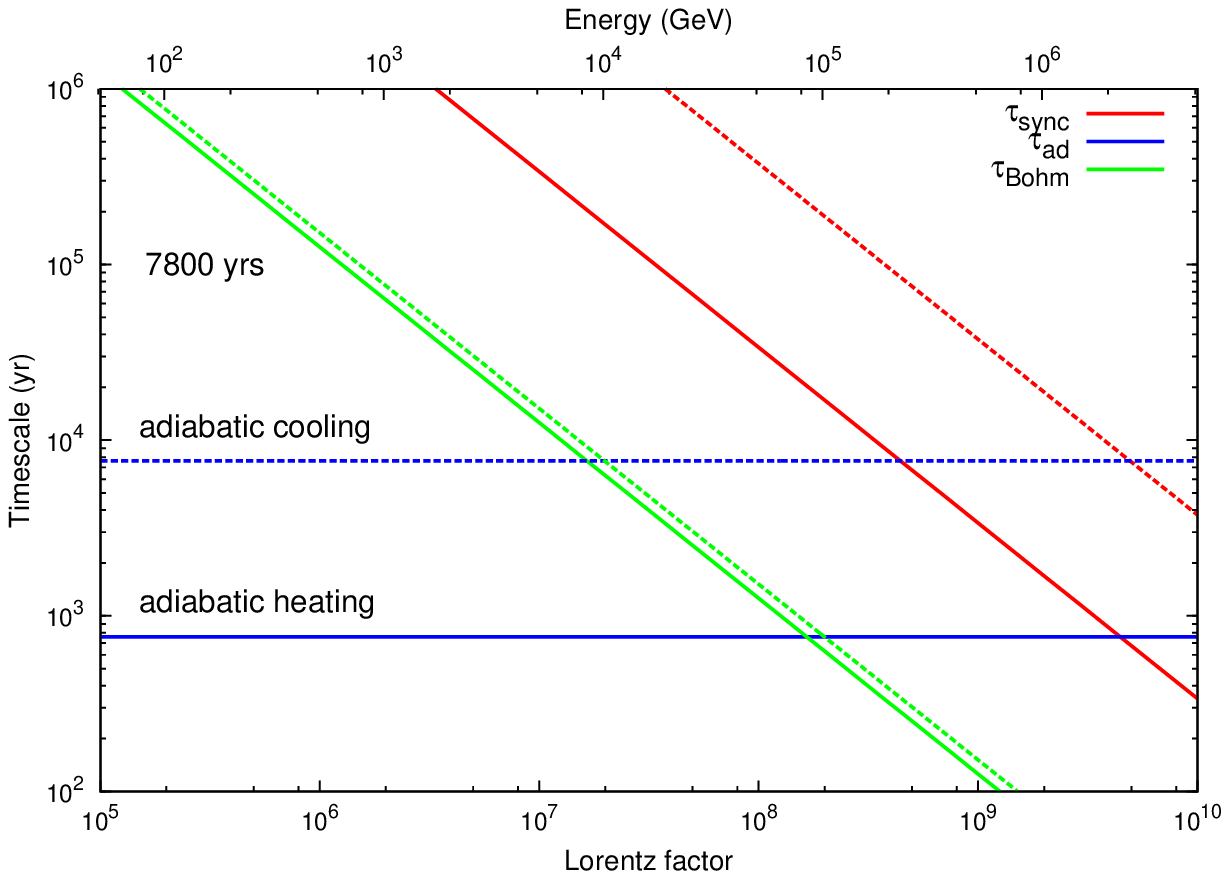}
\includegraphics[scale=0.62]{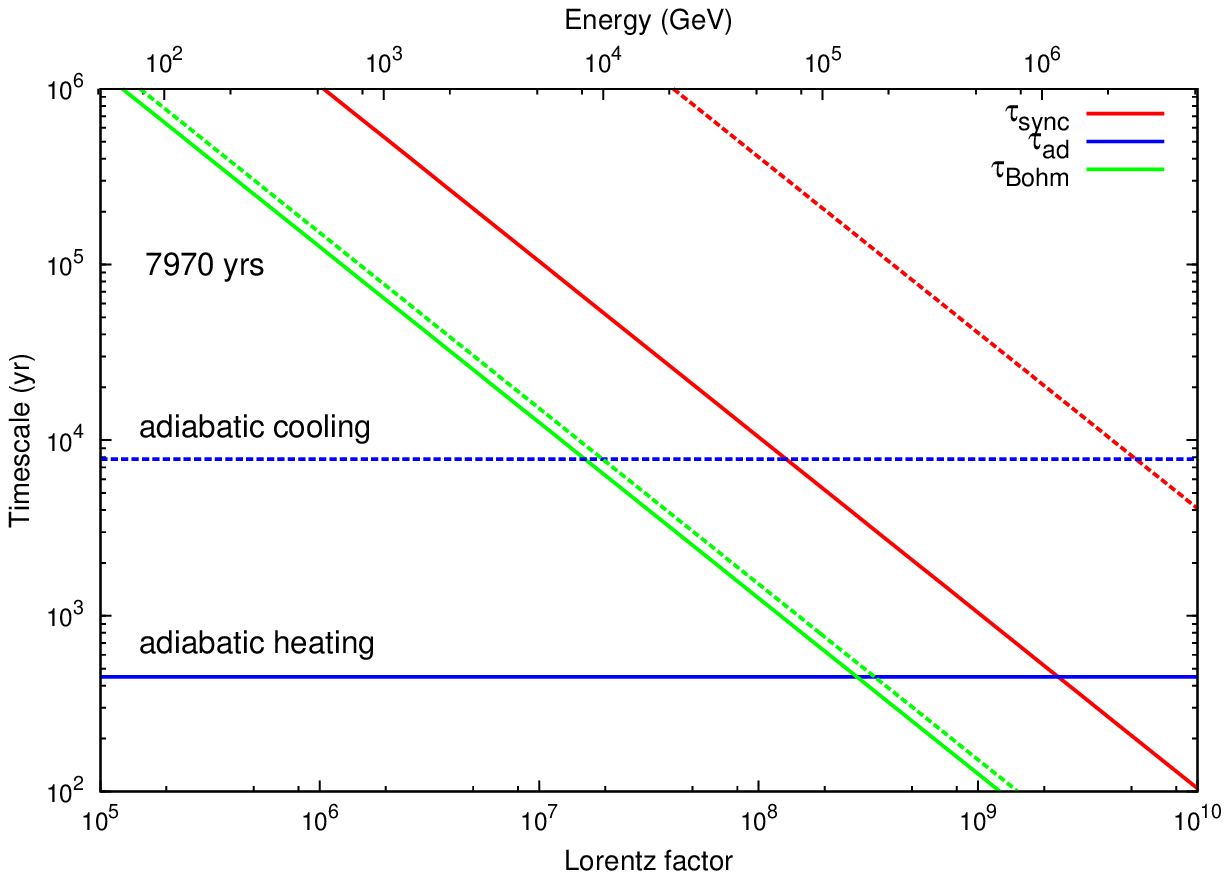}
\caption{Comparison of timescales for relevant particle losses (and energy gain in the case of the adiabatic timescale along reverberation) for models of age and parameters equal  to the matching one shown in Fig. 
\ref{7900-fit-SED}. Shown are models with dynamical evolution without (dashed lines) and with (solid lines) reverberation being considered. Subdominant bremsstrahlung and inverse Compton timescales are not shown for clarity but also considered in the computation.
}
\label{comp1}
\vspace{0.2cm}
\end{figure}
\end{center}

Fig. \ref{comp1} shows that for the current age, 
the losses in both the reverberating and the non-reverberating models are dominated by adiabatic ones at low energies,
and by escaping particles (assuming Bohm diffusion) at high energies. 
The latter is unusual: in general, synchrotron emission dominates the losses of high energy particles.
For \mag's nebula, the magnetic field is so low, that the synchrotron timescale is quite large in comparison to the escape one. 
This low nebular $B$ is a result of the low spin-down power, and is unrelated 
with the high magnetic field at the pulsar surface.

Since escape losses are catastrophic,  this difference is not minor.
The escape of particles is actually removing electrons from the phase space instead of moving 
them into lower energy bins, and thus keeping them for further production of radiation. 
This is considered in the difussion-loss equation that leads to the electron population (see the Appendix), where the escape term is differently considered from the continuous losses, as usual (Gynzburg \& Syrovatskii 1964).
As soon as the energy of the particles is high enough
for the escape timescale to play a significant role, the spectrum of electrons will thus significantly decrease.

\begin{center}
\begin{figure}
\centering
\includegraphics[scale=0.65]{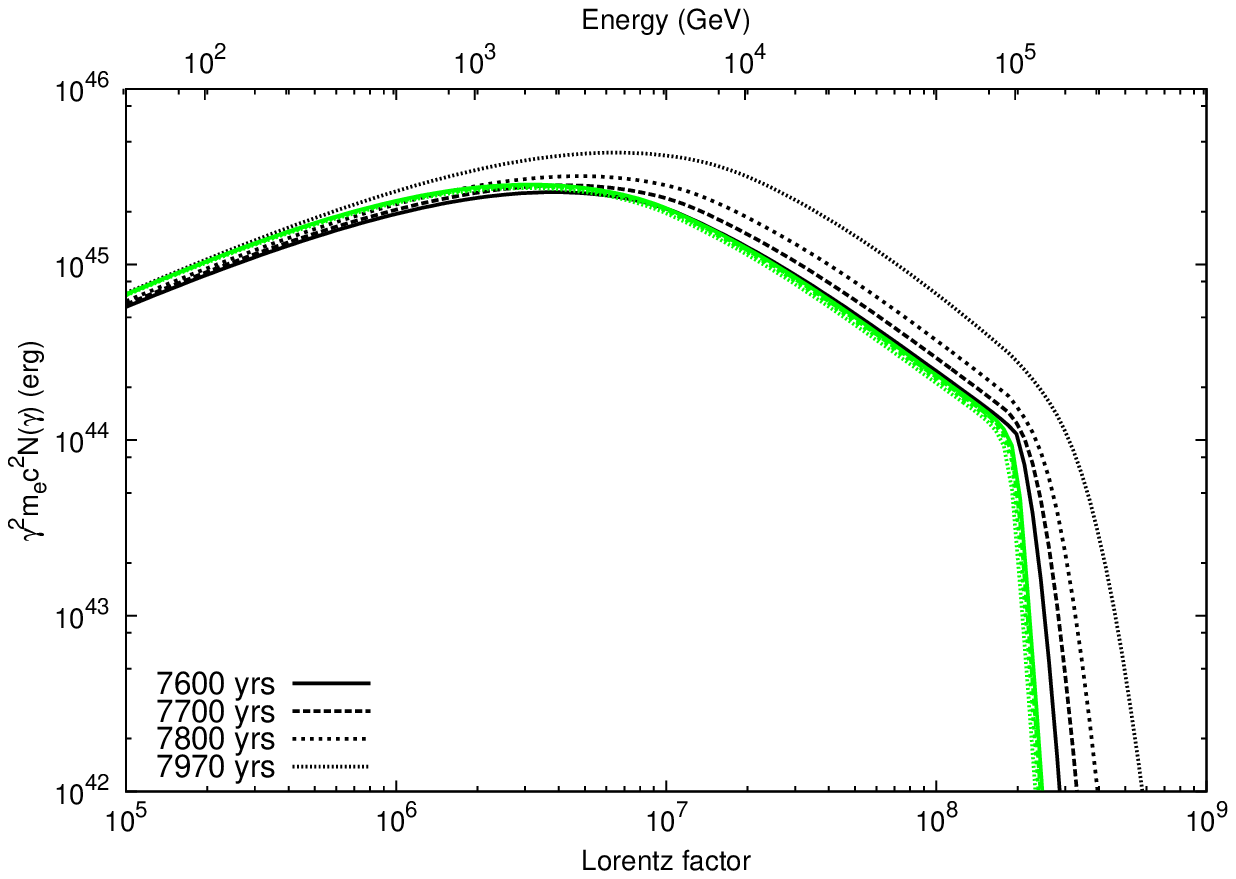}
\includegraphics[scale=0.65]{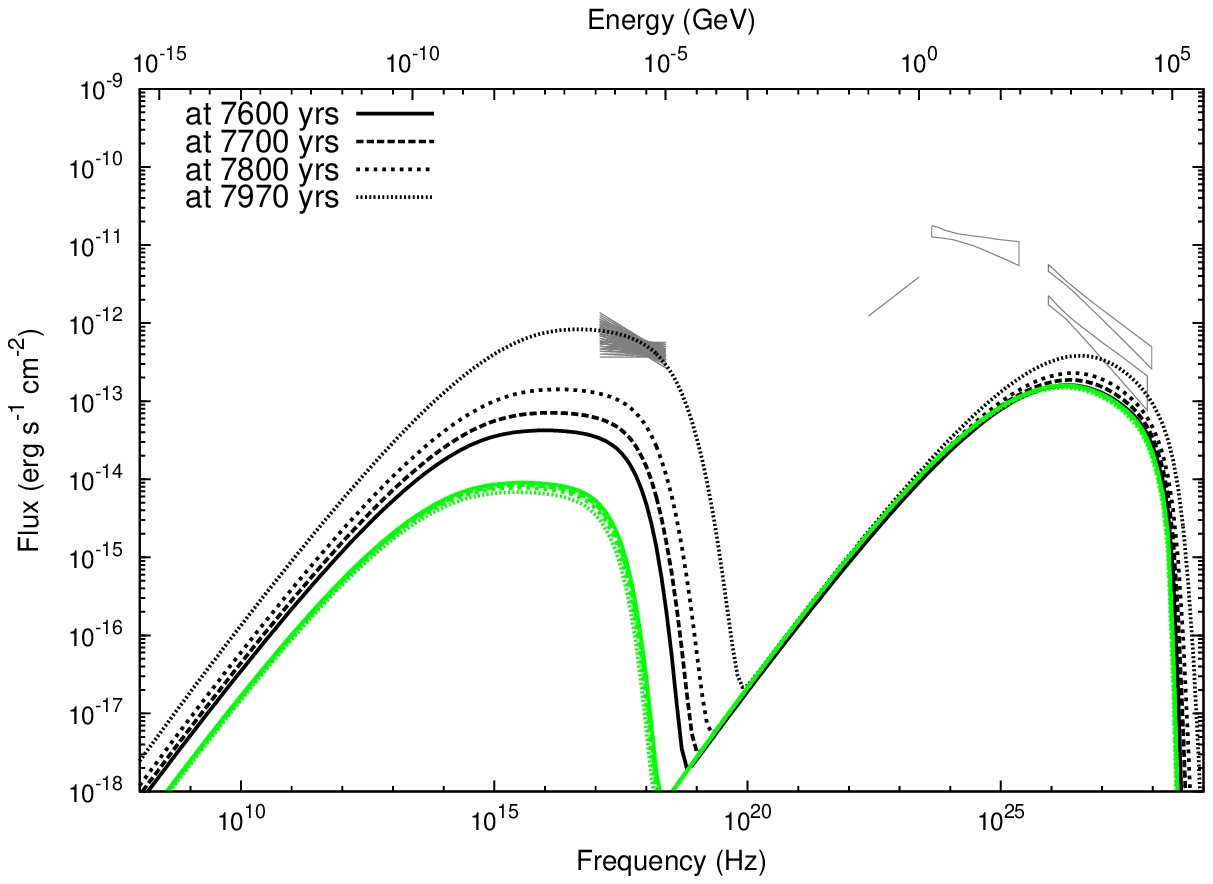}
\caption{Comparison of the time evolution of the electron spectra and SEDs for models of equal age today (7970 yrs) and equal parameters than the matching model shown in Fig. \ref{7900-fit-SED} for a dynamical evolution without (green lines) and with (black lines) reverberation. 
}
\label{comp2}
\vspace{0.2cm}
\end{figure}
\end{center}

The dominance of the escape timescale over the synchrotron one in non-reverberating models 
is even enhanced in comparison with reverberating cases. The magnetic field is lower
(0.7 vs 4.8 $\mu$G at 7970 yrs) and the synchrotron losses are consequently smaller. 
In addition, the radius of the nebula is larger  (8.1 vs 2.9 pc at 7970 yrs)
%as well as the absolute value of the velocity is smaller (also by a factor of a few) 
in the free expanding case, leading to a larger adiabatic timescale.
%so that the timescale for adiabatic losses 
%is larger in comparison with that in the reverberating case. 
%
The latter is independent of the particle's energy, and is essentially unchanged along the evolution 
of the free-expanding nebula, resulting in approximately 8000 yrs. 
This is also the case for the evolution that takes into account reverberation, as long as the compression does not start.
The first panel of Fig. \ref{comp1} shows  the relevant losses at 7000 years 
in both cases considered for the dynamical evolution. Up to that age, the losses, the electron spectrum, and the SED in both models are close to each other.

Subsequent panels in Fig. \ref{comp1} show that whereas the free-expanding evolution keeps an almost constant level,
the adiabatic timescale in reverberation significantly decreases as time goes by. 
Most importantly, the adiabatic timescale along reverberation is no longer representing losses, but energization of particles: the environment is transferring energy to the PWN. 
An smaller adiabatic timescale makes for quick and significant energization of particles that would immediately participate in enhancing the synchrotron spectrum. 
For instance, even at the relevant energies for X-ray production, around $\gamma \sim 1\times 10^8$ and beyond, the 
losses are dominated by diffusion before reverberation, whereas they are rapidly overtaken by the adiabatic heating timescale 
when reverberation is ongoing. Given that the timescale for heating is of the order of the duration of the compression, more and more particles 
participate in generating the X-ray yield.
The dominance of this timescale also translates in the preservation of the overall shape of the electron distribution, and the required (high) energy break of the matching model. The effect upon the energy break is then similar to the case of HESS J1507-622 (see Vorster et al. 2013). 

Without considering this reverberation effect,
the predicted SED would be far from the observational data. 
This can be seen in the second panel of Fig. \ref{comp2}, where a comparison of the time evolution of the SEDs is done. 
For the non-reverberating case, the spectrum does not change its shape, and the only  visible effect is a small reduction of  the nebula particle content (and thus a less luminous nebula) the larger the age. This is expected, since the losses are essentially constant as time goes by, and thus more particles get affected by them the older the nebula is.
Due to this effect, no solution can be found with a nebula in free expansion, not even varying other parameters (e.g., assuming a much larger magnetization). In these models, high energy particles irremediable escape without loosing significant energy, whereas low energy particles are never energized since the adiabatic timescale is always representing a loss. {We also note that non-reverberating models at smaller ages, even at the ages that in this setting would be producing a nebula of the same size than observed  (around 5500 years) will also significantly underpredict the spectrum at X-rays, for the reasons stated. The difference is of several orders of magnitude.}

\subsection{The impact of diffusion}

\begin{center}
\begin{figure}
\centering
\includegraphics[scale=0.65]{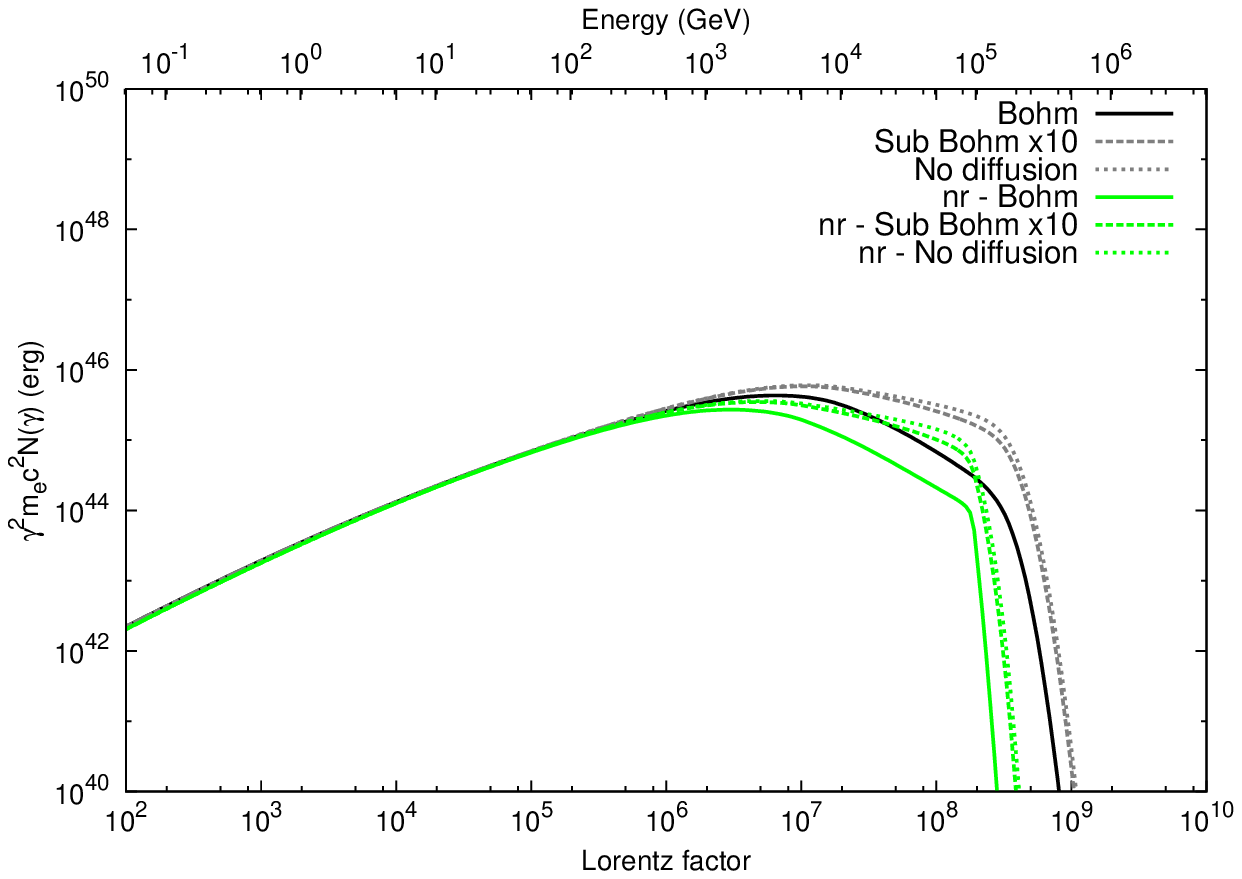}
\includegraphics[scale=0.65]{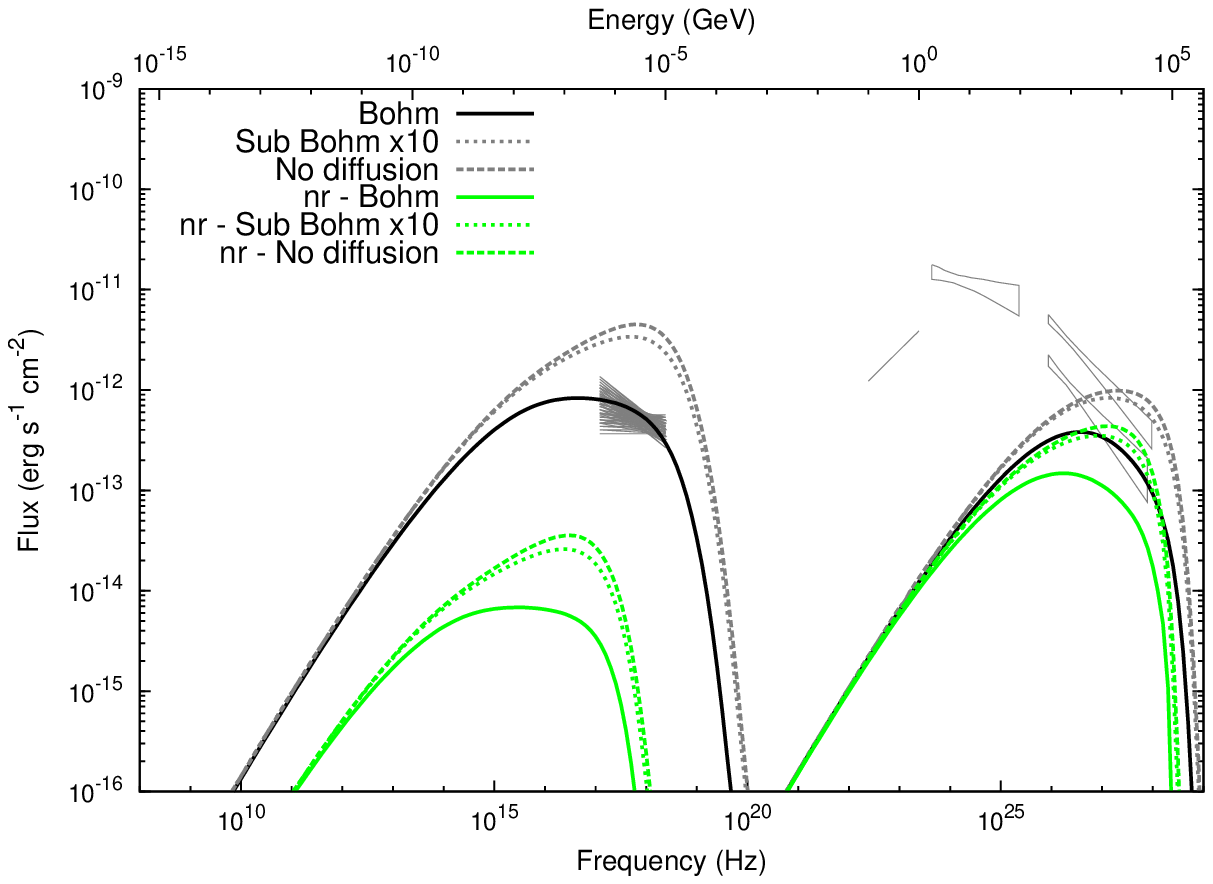} 
\caption{Exploring the impact of diffusion in the solution found. The two panels show: 
%1) Energy of the PWN as a function of time (zoomed for the latest stage of the evolution) for models without diffusion, or with a larger timescale for escape, compared with the matching model of Fig. 1 (black curve) that considers Bohm diffusion. 
1) Influence of the absence or the reduction of the escape timescale in the number of electrons in the nebula. Models with no reverberation (nr) are also shown (for the same set of parameters, in green). The matching model of Fig. 1 is depicted in black. 2) Predicted SED for models in the absence of (or with a reduced) escape. In green we also show the predicted SED for models without reverberation. }
\label{dif}
\vspace{0.2cm}
\end{figure}
\end{center}

 A discussion of the impact of our assumption of the Bohm timescale for diffusion is in order. As stated, diffusion plays an important role in the output of the model since at high particle energies it dominates (over all other processes) the losses of particles (see Fig. \ref{comp1}).  If no diffusion is considered at all (equivalently, if the timescale for diffusion becomes infinity), there will be more particles within the nebula, and thus more energy as well. This effect is clearly visible even if there is sub-Bohm diffusion (where, e.g., the diffusion coefficient is 10 times smaller, and the timescale is 10 times larger than Bohm's).  Fig. \ref{comp1} shows  that if just before reverberation (say at 7000 years) there is a factor of 10 increase in the diffusion timescale, essentially all particles in the nebula (except those at the very high energies, with Lorentz factors beyond $\sim 10^9$) are unaffected by escape. 
The amount of energy lost via diffusion is important (at the start of the reverberation, the difference in energy content in a nebula for which we consider models with and without diffusion can amount up to 50\%). 
These extra energy and particles would yield an important increase in the predicted SED, which would in fact violate the constraints at X-rays and TeV energies by about one order of magnitude. To compensate the predicted SED for that many additional particles there is little one can do, the magnetic field would need to decrease further, and a lower field, in addition, would not correct the unrelated, overpredicted inverse Compton yield, which would then require much smaller photon energy densities, also unlikely for such an active region. These results are summarized in Fig. \ref{dif}.
All in all, from a fitting perspective, a sub-Bohm escape is not preferred. On physical grounds, it also seems more reasonable that pulsars of relatively low spin down power, generating a smaller magnetic field in their nebulae, will also produce less magnetic irregularities, so that diffusion would be fast.

In any case, it is interesting to note that if despite the former arguments one considers that there are more particles in the nebula because diffusion is less efficient, this will not provide a good matching model in cases without reverberation. A sub-Bohm diffusion in a freely expanding nebula cannot mimic the 
reverberation heating needed to explain the observations. 
Without reverberation, even without any diffusion, the additional particles in the nebula do not produce an X-ray yield close to the one observed (it is a factor of 100 lower, see Fig. \ref{dif}). [We recall that in addition, the radius of the nebula without reverberation is much larger than observed, and that if we reduce the age to compensate for the radius, the SED mismatch would increase even further since there are less electrons injected].

\subsection{A comment on model degeneracy}

\begin{center}
\begin{figure}
\centering
\includegraphics[scale=0.65]{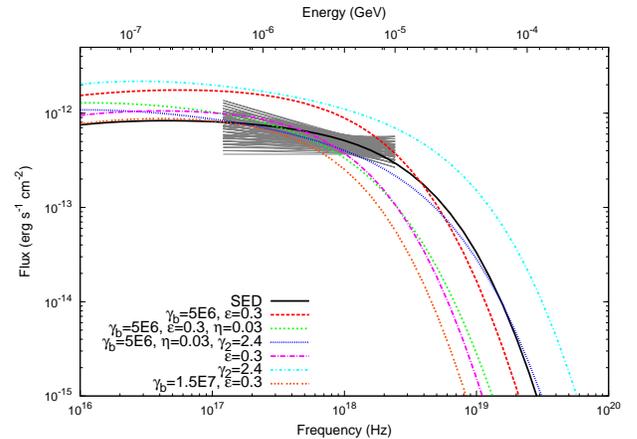} 
\caption{Exploring the degeneracy in choosing matching parameters. All models assume 7970 yrs of age. The black line stands for the model depicted in Fig \ref{7900-fit-SED}.  Only the X-ray part of the spectral energy distribution is shown for clarity. 
The values noted represent the only difference in the parameters of these models in comparison with those shown in the last panel of Table 1.}
\label{deg}
\vspace{0.2cm}
\end{figure}
\end{center}

Without further data (e.g., radio data, or an observationally determined age, or a resolved TeV measurement) we should not conclude on any of the specific values of the matching model of Fig. 1 as being a precise prediction.
Rather, we should see that they  are determining a plausible range.
Reasonable variations around the assumed values of the mass of the ejecta, the ISM density, and the energy of the explosion could also yield to a solution of similar features. However, due to the fact that the solution is in the compression stage, the spectrum is quite sensitive to variations. 
To verify such statement we have run a range of models (all of equal age) varying the low and high injection indices ($\gamma_1=1;1.5$, $\gamma_2=2.1;2.4$),
the energy break ($\gamma_b=5\times 10^6;1\times 10^7;1.5\times 10^7$), the containment factor ($\epsilon=0.3;0.6$), and the magnetization ($\eta=0.03;0.045;0.06$). Not many models are close to the X-ray data. Most of the models overpredict the data either because the population of electrons is too energetic or the magnetic field is too high or both.
Fig. \ref{deg} shows some examples of those models that do not overpredict nor underpredict the data significantly. 

Because of the size constraint discussed in \S 5.1, and under the assumption of a smooth spin-down evolution, there is also not much freedom in choosing the matching age. 
Nebulae that today have different ages (and then have different spin-down age and initial spin-down power) are hard to match to the observational constraint of the size of the nebula, even when some could be close to the spectral energy distribution.

\section{Discussion}

Our simulations show that a rotationally-powered PWN can generate 
the extended emission observed around \mag\ with no other source of energy beyond the pulsar's spin-down power. 
The latter is always (by model
construction) an upper limit to the injected energy in particles, which is given by $(1-\eta)L_{sd}(t)$ at any given time, with $\eta<1$ being
the instantaneous sharing, also known as magnetic fraction. 
Our results do not preclude nor rule out further injection of pairs. 
For instance, more particles could be injected by a yet-unclear transfer of the magnetar's magnetic field, twisted in the inner magnetosphere, into particle acceleration, or by short bursts.
Instead, these results show that an additional source of energy is not needed in order to understand the observations, under the assumption of a smooth spin-down braking.
The latter is the usual assumption for all pulsars and PWNe, and is equally caveated: the pulsar
evolution could be far from such a pleasant ride. How much a magnetar burst, for instance, may change the average properties
of a surrounding PWN, are subjects for future studies.

Here, our study demonstrates the plausibility of a PWN origin of the emission
without invoking any special relationship with the high magnetic field at the surface of the neutron star. 
This is in line with what we have already seen in a systematic study of all
TeV-detected PWNe: Nebula parameters such as field, magnetization, break energy, etc.
neither correlate nor anti-correlate with the pulsar's features, as determined from the spin measurements (Torres et al. 2014).
We have also seen that high-$B$ pulsars can maintain PWNe as well,
and  that efficiencies are not a good tracer of PWN observability 
as soon as a complex time evolution for the particle injection and their burning is considered.

A detailed dynamical-radiative model was used to see that the size and the spectrum of \mag's nebula
can be matched within a rotationally-powered scenario. The requirement for this to happen 
is that the nebula should currently 
be compressed by the environment. 
We have found that this is possible for an age of around 8000 years, about 1.6 times the estimated characteristic age. 
Thus, \mag\ is unrelated with W41 in this model, what seems likely given the existence of several other candidates for such a connection.
The values of the magnetic field and the instantaneous sharing of energy in the nebula
are similar to those found in simulations 
for other systems. \mag's nebula would also be strongly particle-dominated, as all others. 
Perhaps the most notable deviation in resulting parameters appears for the break energy (10$^7$, with the typical value being about few 10$^6$).
Physically, this difference may simply point to the influence of reverberation, through adiabatic heating.

We have shown that diffusion losses (i.e., particles escaping from the nebula) are actually more important at high energies than synchrotron losses. 
This is unusual, and is a result of the low magnetic field that we found in the nebula. This may sound counter-intuitive
taken into account the high field at the surface of the magnetar, but it is not: The low PWN field is a result of the low spin-down and period, and the large size of the nebula, and it is driven in our model in the same way we believe it is for all other  PWNe. 
 We have shown too that a sub-Bohm diffusion is not preferred, and that the absence of diffusion in freely-expanding nebulae cannot produce a good match to the observational data. 
In fact, we have found, that a rotationally-powered PWN can explain the observations only when simulations consider 
reverberation effects. 
In this case, simulations take into account that 
particles can {\it gain} energy via adiabatic heating, thus enhancing the population of pairs that can emit
at X-ray energies.
Particularly, we found that 
the larger is the age along the compression process,  the larger is the energy at which the adiabatic (heating) timescale dominates over the escape one, and the faster the particles gain energy. 
For about a thousand years, then, a high luminosity
can be maintained. As soon as the compression burn off all particles, and the pressure inside the PWN makes it bounce producing a subsequent reduction of the magnetic field, the luminosity
will decrease.
These results are in line (extrapolating several orders of magnitude, towards the magnetar realm) the study of Bandiera (2014). He used a simplified analytical approach to show that middle age PWNe are more likely observable during the reverberation phase. In particular, the spin-down and X-ray luminosity of \mag\ are roughly consistent with the extrapolated curve in figure 3 of Bandiera (2014) (i.e., \mag\ is at or slightly above the extrapolated $t_{c+}$ line of figure 3 in that paper).

The need of reverberation for fitting the spectral and size data
can in fact explain why it would be difficult to find other magnetar nebulae (or in general, other nebulae of pulsars having a  low spin-down power). 
Only those nebulae in the process of reverberating might shine enough to become detectable by our instruments: Before reverberation, the timescales for losses are just too long so that the rate of photon emission is minimal. Too much time after reverberation, all pairs are burnt, the magnetic field is low, and the nebulae may again be invisible.
Since reverberation in such low spin-down power pulsars lasts for about a thousand years, it is reasonable to find few such nebulae. 
The additional effect found so that for a sufficiently low nebular magnetic field the catastrophic losses dominate over synchrotron
adds to the dim character of the systems.

While this work was being prepared, two papers have been published with opposite conclusions to the ones obtained here, i.e., it is impossible to explain the observations of \mag\ with a rotationally-powered nebula (Tong 2016, Granot et al. 2016). While these papers do not present an spectral analysis via simulations as we do, they do present interesting theoretical considerations that apparently imply the inability of \mag\ to rotationally-power the nebula. 
Thus, it is important to discuss our results upon the light of these considerations, in an effort to understand the origin of the divergence in conclusions.

Tong (2016) proposes that the magnetar and the nebula can both be understood in the framework of a wind braking evolution.
Tong's (2016) claim to support the wind braking (and rule out the rotationally-powered) scenario for the nebula is 
that the magnetar's rotational energy loss rate 
is not enough to power the particle luminosity. 
He is considering that only a small portion of the particle energy is converted to non-thermal X-rays, and thus that the particle luminosity of \mag\ should be 10$^{35}$ erg s$^{-1}$ or higher, depending on the X-ray efficiency. 
Since this is beyond the spin-down power, the argument goes, the nebula cannot be rotationally-powered.
This is true only in the case in which there is no accumulation of electrons in the nebula along the lifetime of the pulsar. If we allow for time evolution, and thus for accumulation of all electrons that are not burnt by losses,  one can have an instantaneous income of electrons always limited by the spin-down power at the time of the injection, but many years for accumulating such electrons. 
%
%The numbers that make sense to compare from an evolutionary perspective are then the total spin-down reservoir along the life of the pulsar  
%
%(in erg), the integral of L_sd(t) in time 
%spin down reservoir: 6.6E47 erg
%
%with the accumulated electron population today. 
%In our models, the latter is smaller as can be seen from Fig. \ref{comp2}.
%should be the integral of Fig. \ref{comp2}) 
%
The X-ray emission we see {\it today} should not be directly compared with the electron population {\it injected today} unless their burning is instantaneous, since the emission
may not be  dominated by the fresh electrons, but by the burning of the accumulated pairs. 

Tong's subsequent estimations relies in a number of radiative and dynamical approximations. 
For instance, in an estimate that would be equally applicable to essentially all nebulae of a few pc in size, Tong (2016) imposed a {\it lower limit} to the magnetic field at 240 $\mu$G. This magnetic field is so large for a nebula that is not considered being compressed that not even Crab reaches such value (e.g., Tanaka \& Takahara 2010, Bucciantini et al. 2011, Torres et al. 2013b).
He also considered that the nebula is about a factor of 3 smaller than measured. However, a size of 1 pc ($\sim$50 arcsec) is missing about half the X-ray extended flux.
This is contributing to overestimating the field as well as the needed particle density to achieve the same nebula luminosity. 
He also assumed that the nebula is in equipartition, despite no known nebulae seems to be in such state e.g., see Tanaka \& Takahara (2010), Bucciantini et al. (2011), Torres et al. (2014). 
This yields to a total pressure in particles that is $>200$ times larger than the one we derived in our models. 
% 1 Ba = 1 g cm-1 s-1 = 1 erg cm-3
%
%Finally, the assumed factor of 10 between the PWN radius and the termination shock, which he uses as an argument of consistency for a high particle luminosity, can only be taken as indicative as most. Many of the quoted entries in Table 2 of Kargaltsev \& Pavlov (2008) would have this ratio larger than 10, for instance, it is $>30$ for Kes 75, or $>20$ for G21.5-0.9.
%

Granot et al. (2016)  proposed that the nebula is powered predominantly by outflows from the magnetar, whose main energy source is said to be the decay of the internal magnetic field. 
The conversion mechanism of this internal field into accelerated particles in a wind is not understood.
They have also concluded that in the case \mag\ is related to W41, the magnetar velocity should be at most a few 10 km s$^{-1}$.
They considered an age in the range $5<t_{age}<100$ kyr for the complex, despite the lower end would be contradicting the estimations by Tian et al. referred to above.
By comparing ours with Granot et al.'s work, and despite they seem to echo Tong's argument at times, it would seem that some of the initial assumptions are very similar or the same than those adopted here.
 Among similarities in the approaches are those related to radiation, e.g., our finding of comparable magnetic fields (ours is $\sim$5 $\mu$G, theirs has a fiducial value around  4$\mu$G in a nebula of similar size), or the acceleration constraints considered to fix the maximum energy at injection (although we track this along the time evolution of the nebula since $L_{sd} (t)$, $B(t)$, etc. depend on time).\footnote{Strictly speaking, by measuring a given X-ray energy {\it today} one can only conclude that at some moment in the pulsar's history, there were in the nebula particles energetic enough to produce it.  
An electron of higher energy than those being injected today could have been borne at earlier times, and remain in the nebula for as long as the acting losses over this electron allow.
} Our obtained value of magnetization and their considered range also seems to be comparable.
%(although it is unclear to this reader whether they use equipartition at some point). 
Differences --or at least an unclear direct comparison-- rely on how similar the assumptions for the age and thus the dynamical evolution are. The approaches to deal with reverberation are also different, ours is relying in a direct, numerical  solution of the dynamical set of equations.
We do not find electrons cooling fast by synchrotron emission, but Granot et al. (2016) do not seem to include diffusion losses along most of their analysis. Without the latter, indeed synchrotron losses dominate at high enough energies. We do find a $t_{Synch}$ 
smaller than the estimated age of the SNR W41 (which for us is larger than 50 kyr), and we do not believe \mag\ and W41 are necessarily related.
Other smaller differences may also intervene. For instance, we do not make any radiative approximations in our estimates of synchrotron emission, nor on the determination of the magnetic field along time, nor on the dynamical evolution, nor on the detailed balance (which for us 
is searched by a numerical solution of the full diffusion-loss equation). The concurrency of the impact of all of these approximations is hard to track. 
%We just solve all of this numerically.
%

To finish, we would like to note that \mag\ and the environment of W41, as well as that of \xmmu\ remains an exquisite case for further investigation with forthcoming powerful instruments such as the Square Kilometer (e.g., Taylor 2012)
and the Cherenkov Telescope Arrays (e.g., Actis et al. 2013).
The latter could provide observations sensitive enough to spatially and spectrally separate the contributions to the total TeV emission, thus directly testing not only this work but models of all sources involved. The former could provide constraints to the lower-energy particle population in the nebulae which would help determine model parameters that were currently assumed. We look forward to doing these observations in the near future.

% mencionar otros posibles casos de magnetar nebula? por que esta es la unica.
% comparison with Kes 75 (or rather not if not needed?)

\acknowledgments

We acknowledge support from the grants AYA2015-71042-P, SGR 2014-1073 and the National Natural Science Foundation of
China via grant NSFC 11473027, as well as the CERCA Programme of the Generalitat de Catalunya; and Y. Bao and J. Martin for comments.  

\section*{Appendix: Model formulae}

We show here some definitions used in the models. For details 
see Martin, Torres \& Pedaletti (2016) and references therein. 

\begin{itemize}

\item Spin-down luminosity and age:

\begin{equation}
\label{spindown}
L(t)=L_0 \left(1+\frac{t}{\tau_0} \right)^{-\frac{n+1}{n-1}},
\end{equation}
where $n$ the braking index and where
\begin{equation}
\tau_0 = \frac{2\tau }{ (n-1) } - t_{age}
\end{equation}
and
\begin{equation}
L_0=L_{sd}(1+ \frac{ t_{age}}{\tau_0})^{\frac{n+1}{n-1}}.
\end{equation}

\item Injection of particles:

\begin{equation}
Q(\gamma,t)=Q_0(t)\left \{
\begin{array}{ll}
\left(\frac{\gamma}{\gamma_b} \right)^{-\alpha_1}  & {\rm for \;\;\;}\gamma \le \gamma_b,\\
 \left(\frac{\gamma}{\gamma_b} \right)^{-\alpha_2} & {\rm for \;\;\;}\gamma > \gamma_b,
\end{array}  \right .
\end{equation}

where the normalization term $Q_0(t)$ is computed using the spin-down luminosity of the pulsar $L_{sd}(t)$
\begin{equation}
(1-\eta)L_{sd}(t)=\int_{\gamma_{min}}^{\gamma_{max}} \gamma m_e c^2 Q(\gamma,t) \mathrm{d}\gamma,
\label{eqeta}
\end{equation}

$\eta$ is  the instantaneous sharing parameter, describing the
distribution of the spin-down power into the nebula components.  
It is also 
usually called  the magnetic fraction of the nebula.
No other source of energy is considered.

\item Maximum energy of particles:

The maximum energy that can be achieved is determined either by the synchrotron
limit, established by the balance between the energy loss by particles due to synchrotron radiation 
and the energy gain during acceleration,
\begin{equation}
\label{gsync}
\gamma^{sync}_{max}(t)=\frac{3 m_e c^2}{4 e} \sqrt{\frac{\pi}{e B(t)}},
\end{equation}
%
%where $e$ is the electron charge and $B$ the mean magnetic field of the PWN. 
or by  demanding confinement of the
particles inside the termination shock of the PWN,
%where particles are accelerated implies
%
\begin{equation}
\label{ggyro}
\gamma^{gyro}_{max}(t)=\frac{\epsilon e \kappa}{m_e c^2} \sqrt{\frac{\eta L(t)}{c}},
\end{equation}
being $\kappa$ the magnetic compression ratio (we fix it as 3) and $\epsilon$ is the containment fraction ($\epsilon < 1$).
%This is also the reason why a change in confinement factor affects the electron population, and ultimately the SED.

\item Magnetic field evolution:
\begin{equation}
\frac{d W_B(t)}{dt}=\eta L_{sd}(t)-\frac{W_B(t)}{R(t)} \frac{d R(t)}{dt},
\end{equation}
where $W_B=B^2 R^3/6$ the total magnetic energy, and from where results
\begin{equation}
\label{bfield}
B(t)=\frac{1}{R^2(t)}\sqrt{6 \eta \int_0^t L_{sd}(t') R(t') \mathrm{d}t'}
\end{equation}

\item Diffusion-loss equation:

\begin{equation}
\label{diffloss}
\frac{\partial N(\gamma,t)}{\partial t}=Q(\gamma,t)-\frac{\partial}{\partial \gamma}\left[\dot{\gamma}(\gamma , t)N(\gamma,t) \right]-\frac{N(\gamma,t)}{\tau(\gamma,t)},
\end{equation}

The second 
term on the right hand side takes into account the energy losses due to synchrotron, inverse Compton, and Bremsstrahlung
interactions, as well as the adiabatic losses. The last term on the right hand side accounts for 
escaping particles (we assume Bohm diffusion).

\item Dynamical evolution:

To determine the nebula dynamical state 
we numerically solve the array of equations:
\begin{eqnarray}
\label{radevol}
&& \frac{d R(t)}{dt}=v(t),\\
&& M(t)\frac{d v(t)}{dt}=4 \pi R^2(t) \left[P(t)-P_{ej}(R,t) \right],\\
&& \frac{d M(t)}{dt}=4 \pi R^2(t) \rho_{ej}(R,t) (v(t)-v_{ej}(R,t)) \label{eqmass}.
\end{eqnarray}
The meaning of the variables is given in the text. Eq. (\ref{eqmass}) only applies if $v_{ej}(R,t)<v(t)$. Otherwise, $dM(t)/dt=0$. 

To compute the pressure of the nebula the model considers both the field and the particles, 
such that 
\begin{equation}
P(t)=P_p(t)+P_B(t),
\end{equation}
and where 
\begin{equation}
\label{pmag}
P_B(t)=\frac{B^2(t)}{8 \pi},
\end{equation}
and the pressure contributed by particles is
\begin{equation}
\label{ppar}
P_p(t)=\frac{3(\gamma_{pwn}-1)E_p(t)}{4 \pi R(t)^3}.
\end{equation}
The energy of particles is computed directly from the total population (i.e., it consider radiative losses)
\begin{equation}
\label{epar}
E_p(t)=\int_{\gamma_{min}}^{\gamma_{max}} \gamma m_e c^2 N(\gamma,t) \mathrm{d}\gamma.
\end{equation}
\end{itemize}

%For the density and velocity profiles of assumed ejecta see the main text.
The PWN bounces and starts the Sedov phase when its pressure reaches the pressure of the SNR's Sedov solution, and 
from there onwards it evolves following this equation
\begin{equation}
\label{sedov}
R^4(t_{Sedov})P(t_{Sedov})=R^4(t)P(t).
\end{equation}
We assume $\omega = 9$ for the index of the SNR density power law
as in Chevalier \& Fransson (1992); Blondin, Chevalier \& Frierson (2001); Gelfand, Slane \& Zhang (2009). 
%for a type II SN. 
$v_{ej}$, $\rho_{ej}$ and $P_{ej}$ are the
values of the velocity, density, and pressure of the SNR ejecta at the position of the PWN shell.
They are different when the PWN shell is surrounded  by unshocked ejecta (i.e., the radius
of the PWN is smaller than the radius of the reverse shock of the SNR, $R < R_{rs}$), or by shocked ejecta (where $R_{rs} < R < R_{snr}$,
being $R_{snr}$ the radius of the SNR). The initial profiles for the unshocked medium are assumed following Truelove \& McKee (1999), and Blondin, Chevalier \& Frierson (2001).
To obtain the $v_{ej}$, $\rho_{ej}$ and $P_{ej}$ profiles for the shocked medium, we use the prescription given by Bandiera (1984).
For the shock trajectories we use the semi-analytic model by Truelove \& McKee (1999) for a non-radiative SNR.

\end{document}